\documentclass[review,1p]{elsarticle}

\usepackage{lineno,hyperref}
\usepackage[misc]{ifsym}
\usepackage{amsmath,amssymb,amsfonts}
\usepackage{algorithm2e}
\usepackage{setspace}
\usepackage{graphicx}
\usepackage{textcomp}
\usepackage{xcolor}
\usepackage{diagbox}
\usepackage{multirow}
\usepackage{subfigure}
\PassOptionsToPackage{hyphens}{url}
\usepackage{geometry}
\journal{Journal of Information Processing and Management}

\begin{document}
\begin{frontmatter}

\title{XBlock-EOS: Extracting and Exploring Blockchain Data From EOSIO}
% \tnotetext[mytitlenote]{This is a footnote associated with the title.}

\author[mymainaddress,mysecondaryaddress]{Weilin Zheng}
\ead{zhengwlin@mail2.sysu.edu.cn}

\author[mymainaddress,mysecondaryaddress]{Zibin Zheng\corref{mycorrespondingauthor}}
\cortext[mycorrespondingauthor]{This is to indicate the corresponding author.}
\ead{zhzibin@mail.sysu.edu.cn}

\author[mythirdaddress]{Hong-Ning Dai}
\ead{hndai@ieee.org}

\author[mymainaddress,mysecondaryaddress]{Xu Chen}
\ead{chenx397@mail2.sysu.edu.cn}

\author[mymainaddress,mysecondaryaddress]{Peilin Zheng}
\ead{zhengpl3@mail2.sysu.edu.cn}

\address[mymainaddress]{School of Data and Computer Science, Sun Yat-sen University, Guangzhou, China}
\address[mysecondaryaddress]{National Engineering Research Center of Digital Life, Sun Yat-sen University, Guangzhou, China}
\address[mythirdaddress]{Faculty of Information Technology, Macau University of Science and Technology, Macau}

% \author[mymainaddress,mysecondaryaddress]{Global Customer Service\corref{mycorrespondingauthor}}
% \cortext[mycorrespondingauthor]{This is to indicate the corresponding author.}

% \address[mymainaddress]{1600 John F Kennedy Boulevard, Philadelphia}
% \address[mysecondaryaddress]{360 Park Avenue South, New York}

\begin{abstract}
Blockchain-based cryptocurrencies and applications have flourished in blockchain research community. Massive data generated from diverse blockchain systems bring not only huge business values but also technological challenges in data analytics of heterogeneous blockchain data. Different from Bitcoin and Ethereum, EOSIO has richer diversity and a higher volume of blockchain data due to its unique architectural design in resource management, consensus scheme and high throughput. Despite its popularity (e.g., 89,800,000 blocks generated till November 14, 2019 since its launch on June 8, 2018), few studies have been made on data analysis of EOSIO. To fill this gap, we collect and process the up-to-date on-chain data from EOSIO. We name these well-processed EOSIO datasets as XBlock-EOS, which consists of 7 well-processed datasets: 1) Block, Transaction and Action, 2) Internal and External EOS Transfer Action, 3) Contract Information, 4) Contract Invocation, 5) Token Action, 6) Account Creation, 7) Resource Management. It is challenging to process and analyze a high volume of raw EOSIO data and establish the mapping from original raw data to the well-grained datasets since it requires substantial efforts in extracting various types of data as well as sophisticated knowledge on software engineering and data analytics. Meanwhile, we present statistics and exploration on these datasets. Moreover, we also outline the possible research opportunities based on XBlock-EOS. 
\end{abstract}

\begin{keyword}
Blockchain, EOSIO, Big Data, Data Acquisition, Data Analysis, Security
\end{keyword}

\end{frontmatter}

% \linenumbers

\section{Introduction}\label{sec:introduction}
With the growing prosperity of cryptocurrencies like Bitcoin~\cite{nakamoto2008bitcoin}, blockchain has attracted extensive attention from both academia and industry in recent years. In particular, blockchain can be used in information management systems so as to reduce the dependence of third parties, to improve the interoperability across different business sectors, and to improve efficiency~\cite{BERDIK2021102397,BANIATA2021102393}. Moreover, blockchain can also be used in Internet of Things (IoT)~\cite{WLiang:IoTJ20,ZHAO2020102355,tseng2020blockchain}, critical infrastructures~\cite{YWu:IoTJ20,Alfandi:NOMS20}, enterprise operational management~\cite{PAN2020101946}, anti-fake news~\cite{CHEN2020102370}, vehicle management~\cite{OHAM2021102426}, data management and auditing~\cite{LI2020102382,PUTZ2021102425}.

Blockchain systems can be roughly categorized into permissionless and permissioned blockchains, corresponding to publicly accessing and limited accessing, respectively~\cite{zheng2016blockchain}. Substantial efforts have been made on the permissionless blockchain systems recently, consequently leading to the proliferation of diverse blockchain systems, such as Ethereum~\cite{wood2014ethereum} and EOSIO~\cite{io2017eos}. In a permissionless blockchain system, each peer interacts with a public ledger, which is traceable, tamper-resistant and censorship-resistant. Compared with the traditional Proof of Work (PoW)-based blockchain systems (such as Bitcoin and Ethereum), which are limited by low throughput, EOSIO attempts to offer high throughput with a novel architectural design and Delegated Proof of Stake (DPoS) consensus.

\subsection{Motivation}
Blockchain that serves as a middleware in information management systems as well as other systems (e.g., IoT) generates massive blockchain data. Meanwhile, the prosperous development of permissionless blockchain systems has also led to the generation of massive data. For example, the volume of Bitcoin data has reached 268GB on March 19, 2020 according to the statistics of \textit{BlockChair}\footnote{\url{https://blockchair.com/bitcoin}}. Meanwhile, according to \textit{Etherscan}\footnote{\url{https://etherscan.io/}}, there are more than 16,000,000 smart contracts (including about 230,000 ERC20 token contracts) deployed in Ethereum. Regarding EOSIO, the number of transactions of EOSIO has reached 2.8 billion according to the statistics of \textit{eosflare.io}\footnote{\url{https://eosflare.io/}}, far exceeding those of Bitcoin and Ethereum. Moreover, the architectural design of EOSIO is significantly different from that of Bitcoin and Ethereum, in aspects of the resource management model and consensus mechanism~\cite{lee2019spent}. EOSIO can essentially provide researchers with more diverse types of data than Bitcoin and Ethereum. However, there are few studies on EOSIO data, thereby motivating the study of this paper.

The massive data on blockchain systems has brought huge business values and great opportunities to researchers due to openness, decentralization, and tamper-resistance~\cite{hndai:blockchain-iot2019, zheng2017overview}. Data analysis of massive blockchain data can extract useful information and identify system (or enterprise) bottlenecks, consequently making right decisions for enterprises. In the past, because of security, privacy, and ownership concerns, the real business trading data is usually not opened to researchers; this closure of business data severely hampers related research efforts. In contrast, the data on permissionless blockchain systems are all publicly available to anyone. Meanwhile, the blockchain data can be accessed almost everywhere through the interconnected peer-to-peer (p2p) network. In addition to the trading (transaction) data, many blockchain systems, like Ethereum and EOSIO, also contain both smart contracts and cryptocurrencies (tokens) data. Big data analysis on massive blockchain data cannot only bring huge business values but also promote the development of blockchain. For example, blockchain data analysis can be used for price speculation detection, transaction fraud detection, and smart contract vulnerability detection, consequently improving the security and supervision of blockchains.

At present, the existing studies mainly focus on the data analysis of Bitcoin and Ethereum~\cite{katsiampa2017volatility, wang2019measurement, Weili18, chen2018understanding, tokenscope} while few studies concentrate on EOSIO data analysis. Different from Bitcoin and Ethereum, EOSIO has richer and more diversity of blockchain data mainly due to its unique architecture design in resource management and DPoS consensus scheme. The massive volume of heterogeneous EOSIO data not only brings opportunities but also challenges in data analysis. It is challenging to analyze EOSIO data due to the following difficulties. \textbf{(1) Difficulty in data synchronization.} Although the main network (a.k.a. \textit{mainnet}) of EOSIO has been launched online for less than two years, EOSIO has generated massive volume data mainly due to the high transaction throughput (i.e., a block is generated every 0.5 seconds) thanks to its highly-efficient architectural design. It will take a long time for a newly-joined peer to fully download (or synchronize) the entire EOSIO blockchain data. For example, it takes more than one month and over 500GB storage space to fully synchronize only \textit{entire block data} at a newly-joined peer. Furthermore, it will take longer and require more storage space to  collect other EOSIO data such as the traces and receipts of the transactions. The high requirements (in computing, networking and storage) for data synchronization hinder the efficient analysis of EOSIO data. \textbf{(2) Absence of general data extraction tools for EOSIO.} Although several blockchain websites provide partial (incomplete) EOSIO data, their data extraction tools are generally closed source. Developers cannot design and build their own data extraction tools based on closed source web tools. In addition, these websites generally provide users (even for paid users) with only limited HTTP/HTTPS interfaces to obtain partial EOSIO data. Data acquisition through these websites is slow and incomplete, which seriously impedes the progress of conducting research on EOSIO blockchain. \textbf{(3) Absence of comprehensive data exploration tools for EOSIO.} Although there are a number of studies on data analysis of Bitcoin and Ethereum, including contract security analysis~\cite{luu2016making, kalra2018zeus}, resource management analysis~\cite{chen2017adaptive, pierro2019influence}, there are few studies on EOSIO. As far as we know, only two most recent studies~\cite{huang2020characterizing,quan2019evulhunter} attempted to analyze EOSIO data, while they only analyzed partial EOSIO data (e.g., characterizing the activities in EOSIO~\cite{huang2020characterizing} and detecting fake-transfer vulnerabilities of smart contracts of EOSIO~\cite{quan2019evulhunter}). To the best of our knowledge, there is no work on comprehensive analysis of entire EOSIO data from various data types. \textbf{(4) Difficulties in data extraction and data processing.} EOSIO contains massive heterogeneous data with various types and different data structures (e.g., structural, non-structural data as well as byte code). Moreover, EOSIO has a much higher volume of blockchain data than other representative blockchains (such as Bitcoin and Ethereum). The massive volume of EOSIO data also brings challenges in data processing.

\subsection{Contributions}

To address the above challenges, we introduce a blockchain data analytics framework namely e\underline{X}plore \underline{Block}chain \underline{EOS} (XBlock-EOS) to extract and explore EOSIO data. In particular, we collect raw data consisting of 89,800,000 blocks of EOSIO data from June 8th, 2018 (i.e., the launching date of the EOSIO \textit{mainnet}) to November 14th, 2019. XBlock-EOS contains \textbf{1,882,112 MB} ($\approx$ 1.88 TB) raw data (after compressing JSON format with the highest compression level in zip). The collected raw data includes three types of blockchain data: \textit{blocks}, \textit{transaction receipts}, and \textit{action traces}. Since it is difficult to analyze the massive raw blockchain data, we process and classify the collected EOSIO raw data into seven datasets: \textit{(1) Block, Transaction and Action}, \textit{(2) Internal and External EOS Transfer Action}, \textit{(3) Contract Information}, \textit{(4) Contract Invocation}, \textit{(5) Token Action}, \textit{(6) Account Creation}, \textit{(7) Resource Management}. After processing EOSIO raw data, we obtain well-processed EOSIO datasets with \textbf{203,479 MB} ($\approx$ 198.7 GB, after compressing CSV format with the highest compression level in zip). It is non-trivial to process such a high volume of raw EOSIO data and establish the mapping from original raw datasets to seven well-grained datasets since it requires substantial efforts in extracting various types of data as well as sophisticated knowledge on software engineering and data analytics. We also conduct statistical analysis on the seven well-processed datasets. In addition, we discuss the emerging applications enabled by XBlock-EOS.

In summary, we highlight the major contributions of this paper as follows:
\begin{itemize}
    \item To the best of our knowledge, XBlock-EOS is the first to provide the most comprehensive on-chain well-processed EOSIO data as well as data extraction, statistics and exploration functions to analyze EOSIO blockchain datasets. In contrast to prior studies that only provided partial EOSIO data, XBlock-EOS provides both comprehensive EOSIO raw data and well-processed EOSIO datasets. In particular, XBlock-EOS includes blockchain data, smart contract data, cryptocurrency data, account creation data and resource management data. Moreover, XBlock-EOS\footnote{All the datasets of XBlock-EOS can be downloaded from \url{http://xblock.pro/eosio/}} periodically keeps updating raw datasets as well as processed datasets, all of which have been synchronized with the EOSIO \textit{mainnet}.
    
    \item The XBlock-EOS framework also provides necessary statistics and exploration functions to analyze blockchain datasets. Meanwhile, we also design and develop a new plugin, which can collect EOSIO on-chain data much faster than EOS official plugins. Moreover, we present some statistics and observations from the seven datasets. The well-processed datasets can be easily used for future in-depth data exploration and data analysis. 
      
    \item This paper also outlines the research opportunities brought by XBlock-EOS. In particular, we discuss the applications of XBlock-EOS in aspects of blockchain system analysis, smart contract analysis, and cryptocurrency analysis. Most of these analyses are conductive to blockchain security enhancement. Moreover, the joint analysis of EOSIO data with other blockchain data (such as Ethereum dataset as given in XBlock-ETH~\cite{zheng2019xblock}) can further advance data analysis of blockchain systems and promote the benign development of blockchain.
\end{itemize}

The remainder of this paper is organized as follows. Section~\ref{sec:back} firstly gives an overview of EOSIO, highlighting its differences from other permissionless blockchain systems. Section~\ref{sec:rawdata} then presents raw data acquisition from EOSIO. Section~\ref{sec:dataexploration} next presents a statistical analysis on seven refined datasets. Section~\ref{sec:application} discusses the applications of XBlock-EOS data and Section~\ref{sec:related} investigates the related work about XBlock-EOS. Finally, we provide a summary of this paper in Section~\ref{sec:conc}.

\section{Background}\label{sec:back}
\begin{figure}
    \centering
    \includegraphics[width=4.4in]{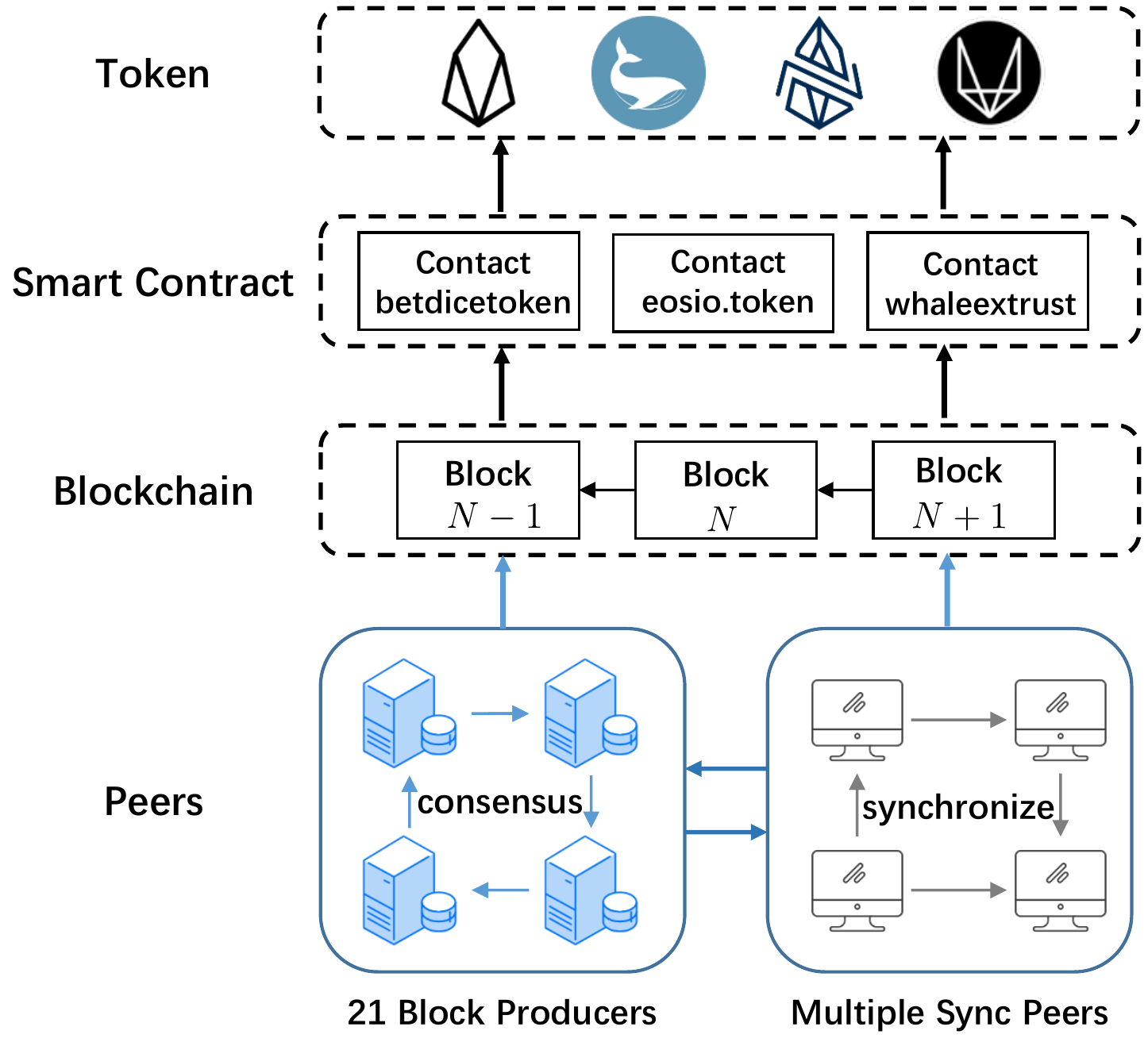}
    \caption{Overview of EOSIO Blockchain}
    \label{0_arc}
\vspace*{-0.25cm}
\end{figure}

Figure~\ref{0_arc} presents an overview of EOSIO blockchain, which consists of four layers from bottom to top: \emph{peers}, \emph{blockchain}, \emph{smart contract}, and \emph{token}. We next review the basic concepts in each layer of EOSIO.

\subsection{Peer and blockchain}
In short, a blockchain system is essentially a distributed ledger with a chain-like data structure consisting of a number of connected blocks. Transactions are packaged into blocks, each of which is confirmed by the entire network through a consensus protocol in a period of time. Unlike Bitcoin and Ethereum, EOSIO uses Delegated Proof-of-Stake~\cite{larimer2014delegated} instead of Proof-of-Work or Proof-of-Stake as its consensus mechanism. In DPoS of EOSIO, only 21 block producers (consensus peers) can produce blocks and verify transactions while other sync peers only synchronize the blockchain data. In contrast, any miners in Ethereum and Bitcoin have the opportunity to undertake this work. Once the new block has been confirmed by most peers, i.e., $14$ $(=2/3 \times 21)$ consensus peers in the EOSIO network, it is considered as completed. In other words, the EOSIO blockchain enhances the reliability of transaction data by copying calculation and storage across multiple peers.

Due to the integrity of the blockchain data in each permissionless peer, researchers can obtain the entire blockchain data by connecting a new peer to the blockchain network. Blockchain data essentially save all operations performed by real-world users on the blockchain network, thereby containing substantial business value. For example, a transaction is essentially an operation performed by different business parties. Therefore, a transaction may imply a potential interest relationship between any two people or entities. Big data analysis on blockchain can help understand user behavior in real-world economic systems. Moreover, the rapid technical development of blockchain systems has boosted the growing number of blockchain users as well as transactions, leading to a massive growth of blockchain data. In particular, EOSIO that generates a block every 0.5 seconds on average has much higher transaction throughput per second (tps) than Bitcoin and Ethereum. Therefore, the data growth rate of EOSIO is much higher than that of Bitcoin and Ethereum. The analysis on such massive data is challenging while it also brings huge business values via EOSIO data analysis.

\subsection{Smart contract}
Smart contract, as a promising technology aimed to reshape the modern industry, was proposed earlier than blockchain~\cite{szabo1997idea}. However, it was not well developed until the advent of Ethereum (i.e., the first Turing-complete blockchain smart contract platform), in which smart contract really plays its role of assuring trustworthy transactions between any two parties without a third party's intervention. Blockchain-based smart contracts are essentially computer programs, in which execution states are stored on blockchain. The blockchain transactions represent the deployment or invocation of smart contracts, triggering updates to the state of blockchain. Blockchain guarantees the reliability of smart contracts by replicating the computation of smart contracts at the peers in the network.

Currently, most blockchain systems have enabled smart contracts. For example, the Bitcoin system enables users to run a simple script program during transaction execution, which can be regarded as one of the simplest smart contracts. However, Bitcoin scripts that are not Turing-complete cannot support complex logic. In contrast, the prosperous blockchain systems, such as Ethereum and EOSIO, can well support Turing-complete smart contracts. Smart contracts run in an environment called a blockchain virtual machine. In particular, in Ethereum, smart contracts run in the Ethereum virtual machine (EVM), while EOSIO smart contracts run in the WebAssembly-based EOS virtual machine (EOSVM). In order to solve the \textit{halting problem}, Ethereum introduced a \emph{gas mechanism} to prevent the malicious behavior of a smart contract, such as an infinite loop. Miners in Ethereum use \textit{Gas} as a unit to measure the computation of each operation (instruction) of the smart contract since \textit{Gas} is a scarce resource that is acquired by cryptocurrency purchase. 

Unlike many public blockchain systems with gas mechanism, EOSIO solves the \textit{halting problem} by limiting the RAM, CPU, and Network (NET) resources of smart contracts. All these three resources can be obtained only after users mortgage some EOS. Among them, RAM is used to limit the storage of the contract while it is an unrecoverable resource. In other words, the consumption of RAM only depends on how much data is stored. Contracts (or users) can delete the stored data to reclaim RAM, and sell RAM for EOS. CPU and NET limit the computation and the network transmission of the contract, respectively. They are recoverable resources and can be recovered after 24 hours when exhaustion. Users can redeem the mortgaged CPU and NET at any time and receive the corresponding EOS after three days. Consequently, EOSIO is regarded as a free blockchain platform for users, and many DApps in EOSIO are willing to provide users with these resources.

\subsection{Tokens and clients}
Since Ethereum introduced the standard token protocol (also known as a template) in two smart contracts: \textit{ERC20} and \textit{ERC721}, \textit{Initial Coin Offering (ICO)}~\cite{chohan2017initial} has swept the entire cryptocurrency ecosystem. The emergence of ICO has greatly enriched the ecosystem of permissionless blockchain, making blockchain system a more flexible distributed financial system. EOSIO has no exception, and a standard token token protocol was introduced at the beginning of its launch. EOS, as one of the most representative tokens in EOSIO, has been used for daily operations, such as transferring, leasing, creating an account, buying RAM, staking CPU, and staking NET. In addition, as shown in the top layer of Figure~\ref{0_arc}, there are three other well-known tokens \emph{WAL}, \emph{NUT}, \emph{VTX}. Everyone can publish a standard token contract in EOSIO to create and issue tokens. After that, any other users and smart contracts can send or receive tokens without a third-party. In Section~\ref{Dataset 5: Token Action}, we explore the data of tokens in EOSIO.

EOSIO allows any computers that meet the requirement of the protocol like p2p protocols to join the network. The official EOSIO development team provides an EOSIO client called \texttt{\small Nodeos}. Anyone using \texttt{\small Nodeos} can join the network. \texttt{\small Nodeos} provides a standard JSON-RPC interface through \texttt{\small eosio::http\_plugin}\footnote{\url{https://eosio.github.io/eos/latest/nodeos/plugins/index}\label{plugin}} and \texttt{\small eosio::chain\_plugin}\textsuperscript{\ref{plugin}} for users to interact with the EOSIO blockchain. Moreover, the official EOSIO development team also develops a tool named \texttt{\small Cleos}, which is a command-line tool that interfaces with the API exposed by \texttt{\small Nodeos}. Through the interfaces and tools, users can obtain block data from EOSIO. However, in order to obtain the receipts and traces data, it is necessary to replay all transactions and store the generated receipts and traces in memory through \texttt{\small history\_plugin}\textsuperscript{\ref{plugin}}. Since EOSIO produces one block every 0.5 seconds, its total transaction volume is much higher than that of Ethereum. Replaying such a large number of transactions takes a long time and consumes a lot of memory. EOSIO development team develops \texttt{\small state\_history\_plugin}\textsuperscript{\ref{plugin}} and \texttt{\small mongo\_db\_plugin}\textsuperscript{\ref{plugin}}, which can cache receipts and traces into database engines, such as PostgreSQL and MongoDB. These plugins do not require a lot of memory, but make transaction replay slower because it takes extra time to insert traces and receipts into the database engines. \emph{To overcome these challenges, we developed our own plugin, which is suitable for our data acquisition and exploration}. We will show more details about data acquisition of blockchain data in Section~\ref{sec:rawdata}.

\section{Raw data extraction from EOSIO}
\label{sec:rawdata}
\begin{figure}
    \centering
    \includegraphics[width=4.6in]{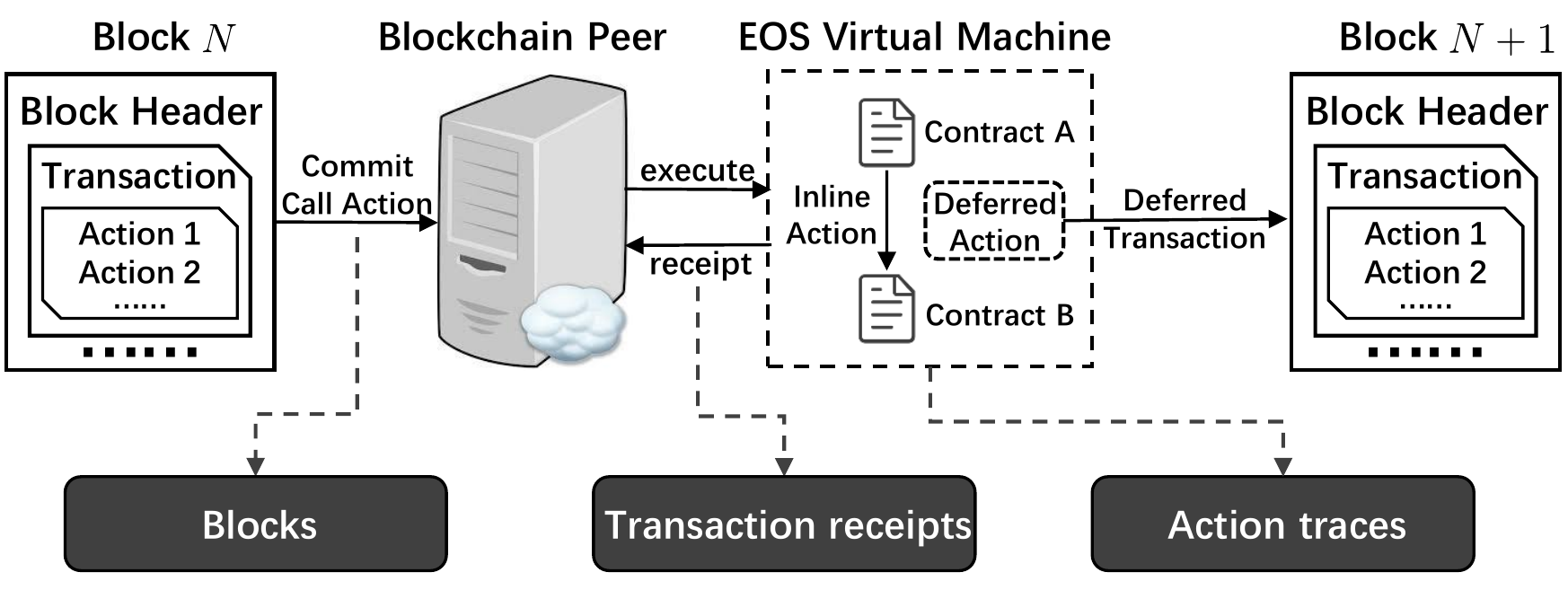}
    \caption{Raw data collection during EOSIO transaction flow}
    \label{0_rawdata}
\vspace*{-0.25cm}
\end{figure}

This section describes the process of obtaining raw data from the EOSIO blockchain. Figure~\ref{0_rawdata} illustrates the typical EOSIO transaction execution flow, from block $N$ to block $N+1$, with the EOSVM of the blockchain peer in the middle. During this procedure, we collect three types of raw blockchain data: Blocks, Transaction receipts, and Action traces. We next describe the details on the composition and acquisition of each kind of raw data.

\subsection{Blocks}
Block data is directly stored in EOSIO blockchain. Each block mainly consists of two elements:
\begin{itemize}
    \item \textbf{Block Header: } Block header consists of some basic information of a block, including the block producer, timestamp, transaction root, etc.
    \item \textbf{Transactions and Actions:} Transactions construct the body of the block and each transaction consists of one or multiple actions. Each action represents a call to a smart contract, mainly including the following fields: account (contract name), name (contract function name), data (function parameter), authorizations (authorizers).
\end{itemize}

It is worth noting that EOSIO's actions can be mainly categorized into three types: \textit{calling action}, \textit{inline action}, and \textit{deferred action}~\cite{xu2018eos}. A \textit{calling action} represents a user's call to a contract and an \textit{inline action} represents a call within the contracts or between the contracts. An \textit{inline action} is generally triggered by a \textit{calling action} and is completed in the same transaction (block). Failure of either \textit{inline action} or \textit{calling action} will cause the transaction to fail. A \textit{deferred action} is used to initiate a deferred transaction (generally packaged in a transaction of a block in the future), and its execution result does not affect the original transaction. It is important to note that only \textit{calling action} and \textit{deferred action} are explicitly packaged into a transaction in a block.

At present, the EOSIO development team provides users with \texttt{\small Nodeos} to synchronize data on the \textit{mainnet} (the main network of EOSIO). There are two major manners to synchronize data: 1) starting \texttt{\small Nodeos} from the genesis block, and 2) downloading the blocks from some EOSIO backup service provider such as EOS Amsterdam\footnote{\url{https://snapshots.eosamsterdam.net/}}, and starting \texttt{\small Nodeos} from the specified block. In order to obtain the data faster, we adopt the second method. By activating the \texttt{\small chain\_plugin} and \texttt{\small http\_plugin}, users can obtain each block through the RPC interface of Nodeos. However, the block data only contains partial information about \textit{calling action} and \textit{deferred action}, which is not enough to comprehensively analyze blockchain users. In addition, through these block data, we cannot obtain the resource consumption of the transaction (e.g., RAM) and the details of the transaction execution (e.g., what errors occurred and which other contracts were called during the transaction execution). 

\subsection{Action traces}
Action trace data is essentially the detailed run-time data of each action that is generated in EOSVM (e.g., calls within or between the contracts, transferring EOS tokens from a contract to others). With action trace data, we can collect the detailed information about the inline actions and deferred actions (transactions). Combined with the information of the block data, we can collect the complete information about a transaction, such as which contracts and which functions are called, whether it is a deferred transaction.

Action trace data cannot be obtained or observed from the block data, but can be recorded during the transaction execution. Therefore, we need to replay all transactions and collect all action traces in this procedure. The EOSIO development team provides  the \texttt{\small history\_plugin} to cache the generated traces in memory. The \texttt{\small history\_plugin} supports extremely fast querying of traces, but it also consumes huge memory due to the huge transaction volume of EOSIO. At present, almost all EOSIO full peers have closed this plugin. To address this problem, the EOSIO development team develops \texttt{\small state\_history\_plugin} and \texttt{\small mongo\_db\_plugin}, which insert the traces into database engine. These plugins aim to reduce the memory requirement and support the convenient and fast query of traces. However, these plugins would also slow down the replay procedure because it takes extra time for the official plugins to insert the traces into the database engines. %Moreover, it may also take extra procedures for machine learning (ML) tools to deal with the database files since most ML tools prefer dealing with tabular files (e.g. csv files). 

\begin{figure}[t]
\centering
\subfigure[EOSIO official plugin]{
\includegraphics[height=4.0cm]{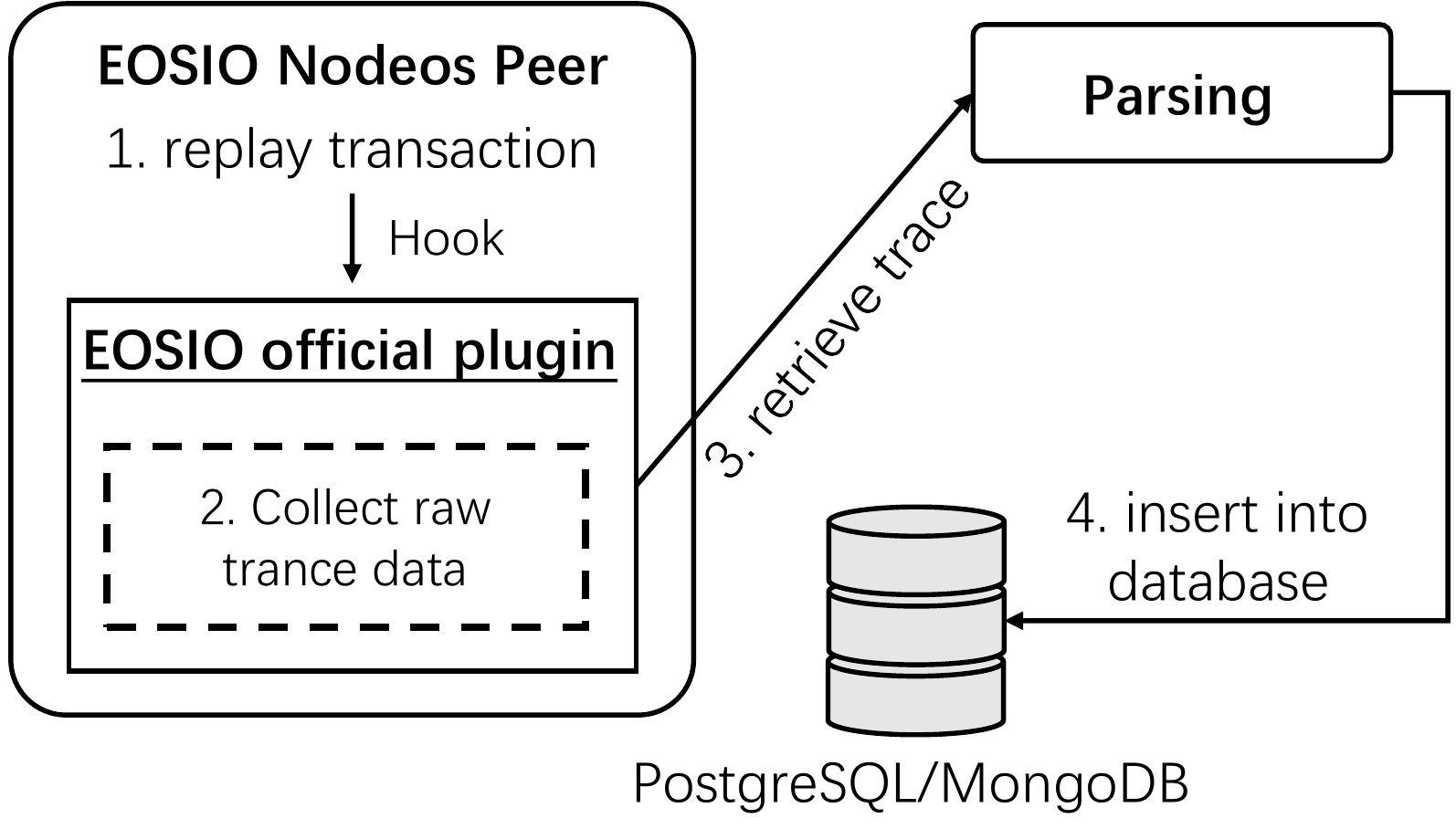}
\label{fig:plugin-a}}\hfil
\subfigure[Our \texttt{history\_file\_plugin}]{
\includegraphics[height=4.0cm]{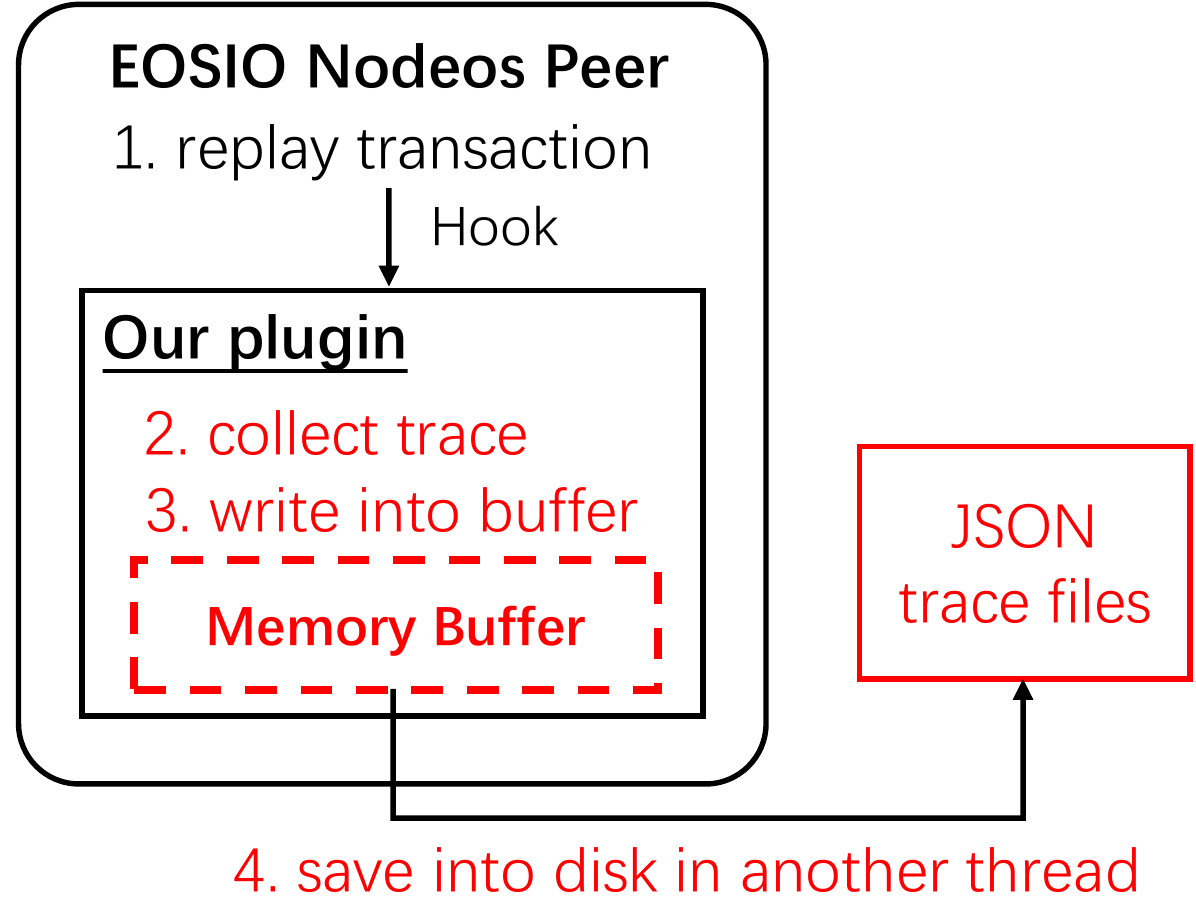}
\label{fig:plugin-b}}\hfil
\caption{Comparison of EOSIO official plugin with our \texttt{history\_file\_plugin}}
\label{fig:eosioplugin}
\vspace*{-0.25cm}
\end{figure}

We first analyze the working flow of EOSIO official plugins. As shown in Figure~\ref{fig:plugin-a}, \texttt{\small Nodeos} will hook \texttt{\small state\_history\_plugin} or \texttt{\small mongo\_db\_plugin} when one of them is activated. These plugins collect the raw trace data when replaying transactions, then retrieve the traces and parse them into the well-formatted data being suitable for some specific database engines (such as PostgreSQL and MongoDB). Finally, the formatted traces are inserted into the database according to certain indexes. However, data parsing and insertion may slow down the replay procedure, which is not conducive to the rapid collection of trace data, especially for massive EOSIO data. Moreover, parsing such huge trace data requires extensive computing and storage resources, thereby resulting in the frequent crash of the \texttt{\small Nodeos} peer. It often requires substantial time, memory, and storage to collect traces with these plugins.

To address this issue, we design and develop a new plugin namely \texttt{\small history\_file\_plugin} to support the rapid collection of EOSIO trace data for subsequent processing and analysis in our XBlock-EOS. Figure~\ref{fig:eosioplugin} shows a comparison between EOSIO official plugin and our \texttt{\small history\_file\_plugin}. As shown in Figure~\ref{fig:plugin-b}, our \texttt{\small history\_file\_plugin} collects raw trace data and writes them into \textit{Memory Buffer} when replaying transactions. Then, another thread asynchronously reads trace data from \textit{Memory Buffer}, serializes, and saves them into storage devices (e.g., HDDs or SSDs) periodically. Our plugin can directly save traces as multiple files in the JSON format, i.e., a semi-structured data type similar to that in MongoDB collected by official plugins. Therefore, its data collection speed is much faster than other plugins. For example, under the same machine with a 12-core Intel Core i7-5820K@3.30GHz processor and 128GB memory, our \texttt{\small history\_file\_plugin} only takes about one day to collect the trace data of the first 20 million blocks while EOSIO official plugins take more than a week for the same amount of trace data. Meanwhile, \texttt{\small history\_file\_plugin} also supports the collection of traces for specific block intervals. Moreover, the collected raw trace data (i.e., JSON format) also contain duplicates and redundant data. To address this challenge, we also implement several scripts to mitigate the redundant information while preserving the necessary information, and consequently save the trace data into tabular files (i.e., CSV format), which are well suitable for most machine learning tools. 

In summary, our \texttt{\small history\_file\_plugin} can better meet the needs of fast data collection for data collectors and facilitate the further data analysis tasks. The modified \texttt{\small Nodeos} source code has also been published on the \textbf{\textit{XBLOCK.PRO}} website\textsuperscript{\ref{website}}.

\subsection{Transaction receipts}
In EOSIO blockchain, transaction receipts are generated after transactions are executed, which can be read by external clients or people, but cannot be obtained by the internal EOSVM. A transaction receipt records the execution of a transaction. More importantly, it contains resource consumption information for a transaction. Since EOSIO's resource management model is significantly different from other blockchains, it is necessary to collect transaction receipts and analyze user (transaction) resource usage. It can help us better understand the user and contract resource usage characteristics in EOSIO and the resource ecology of the EOSIO blockchain.

The collection of transaction receipts is similar to the collection of action traces. By activating \texttt{\small history\_file\_plugin}, \texttt{\small Nodeos} can save the receipts of all transactions in a certain block interval as JSON format files. We then quickly screen out the useful information from these files.

In short, there are three kinds of raw blockchain data that can be obtained from EOSIO: Blocks, Action traces, and Transaction receipts. Our plugin speeds up the collection of raw data and saves it in the JSON format according to the block number, similar to that saved in MongoDB using \texttt{\small mongo\_db\_plugin}. JSON is a semi-structured data format, which nevertheless is not the most common format used by data analytics (e.g., machine learning or data mining tools). Moreover, both EOSIO official plugins and our developed plugin collect EOSIO trace data containing duplicated and redundant information. It is necessary to simplify data representation and fasten data analysis for further research. Therefore, we implement some corresponding scripts to extract the necessary information from the collected trace data (i.e., JSON format) and save them as CSV format files. In particular, we obtain seven well-processed datasets to be elaborated in Section \ref{sec:dataexploration}. The well-processed datasets can be easily analyzed by most data analysis tools. It is worth noting that these scripts can run in sync with our plugin to maximize the collection speed. All scripts have also been published on the \textbf{\textit{XBLOCK.PRO}} website. These tools can be modified and tuned to fulfill different data analytics tasks.

It is well known that EOSIO is enabling millions of transactions per second. So, it is challenging to handle this huge amount of transactions. Although the average transaction throughput of EOSIO \textit{mainnet} is only about 56.42 tps in fact (as shown in Section~\ref{Dataset 1: Block, Transaction and Action}), it is nearly 4 times as high as that of Ethereum and 8 times as high as that of Bitcoin. According to the above analysis, if we apply the official plugins to the raw data collection of XBlock-EOS, it will require higher hardware resources and take longer time to collect EOSIO trace data. In addition, subsequent data processing needs to constantly query from the database engines, thereby significantly affecting the performance of data insertion and slowing down data collection. The official plugins prevent data collection and processing from running simultaneously, thereby greatly limiting the scalability. Our \texttt{\small history\_file\_plugin} not only speeds up the raw data collection, but also separates data collection and data processing. It allows data collection and data processing to be carried out simultaneously, consequently improving the scalability of XBlock-EOS. XBlock-EOS regularly downloads relevant snapshot and backup data from EOSIO backup service providers (e.g., EOS Amsterdam), and collects new data from a specific block number to keep up with the \textit{mainnet}. Thus, our developed plugin as well as other corresponding tools can greatly enhance the scalability of XBlock-EOS to handle the huge amount of transactions on the EOSIO \textit{mainnet}.

\section{Data exploration of EOSIO} 
\label{sec:dataexploration}

\begin{figure}[t]
    \centering
    \includegraphics[width=7.5cm]{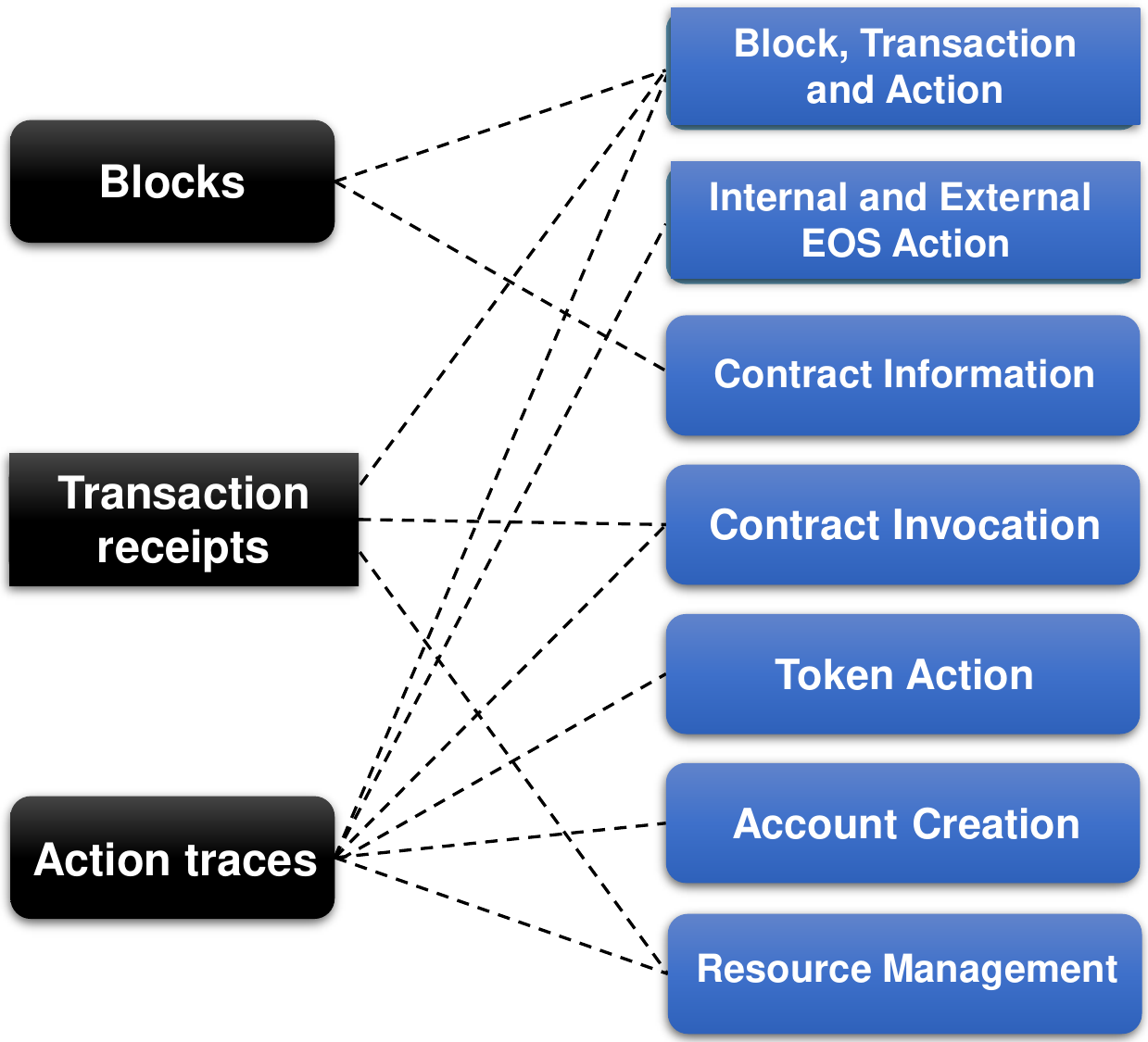}
    \caption{Mapping from raw data to seven datasets}
    \label{0_washdata}
\vspace*{-0.25cm}
\end{figure}

In this section, we process the obtained raw data from EOSIO and divide the raw data into seven datasets: (1) Block, Transaction and Action, (2) Internal and External EOS Transfer Action, (3) Contract Information, (4) Contract Invocation, (5) Token Action, (6) Account Creation, (7) Resource Management. Figure~\ref{0_washdata} shows the categorical relationship from the raw data to the seven datasets. We can observe that \textit{Action traces} are the most widely used in data processing. Next, we will introduce how these seven datasets are generated, and show some statistics and observations about the datasets.

\subsection{Dataset 1: Block, Transaction and Action} \label{Dataset 1: Block, Transaction and Action}

In order to investigate the basic statistic information of EOSIO, we extract the blocks, intra-block transactions and intra-transaction actions from EOSIO. In particular, there are 89,800,000 blocks, 2,533,292,528 transactions and 2,916,530,553 actions (excluding inline actions). We also calculate the average CPU and NET usage of every block according to the \texttt{\small cpu\_usage\_us} and \texttt{\small net\_usage\_words} of the transaction. Moreover, we count the information of block producers to measure the degree of decentralization of EOSIO.

\begin{table}[h]
    \setlength{\belowcaptionskip}{0.2cm}
    \centering
    \caption{Statistics of Dataset 1}
    \renewcommand{\arraystretch}{1.5}
    \footnotesize
    \begin{tabular}{|l|r|}
    \hline
    \textbf{Statistics} &  \textbf{Values} \\ \hline\hline
    No. of Blocks     &   89,800,000 \\ \hline
    No. of Transactions &  2,533,292,528    \\ \hline
    No. of Deferred transactions & 357,455,192 \\ \hline
    No. of Actions (excluding inline actions) & 2,916,530,553 \\ \hline
    No. of Block producers & 63 \\  \hline
    Mean of Transaction Count per Block & 28.21  \\ \hline
    \end{tabular}
    \label{Statistics of Dataset 1}
\vspace*{-0.25cm}
\end{table}

\begin{figure}
\centering 
\subfigure[\scriptsize Word-Cloud Statistics of Block Producers]{
\centering
\includegraphics[width=5.3cm]{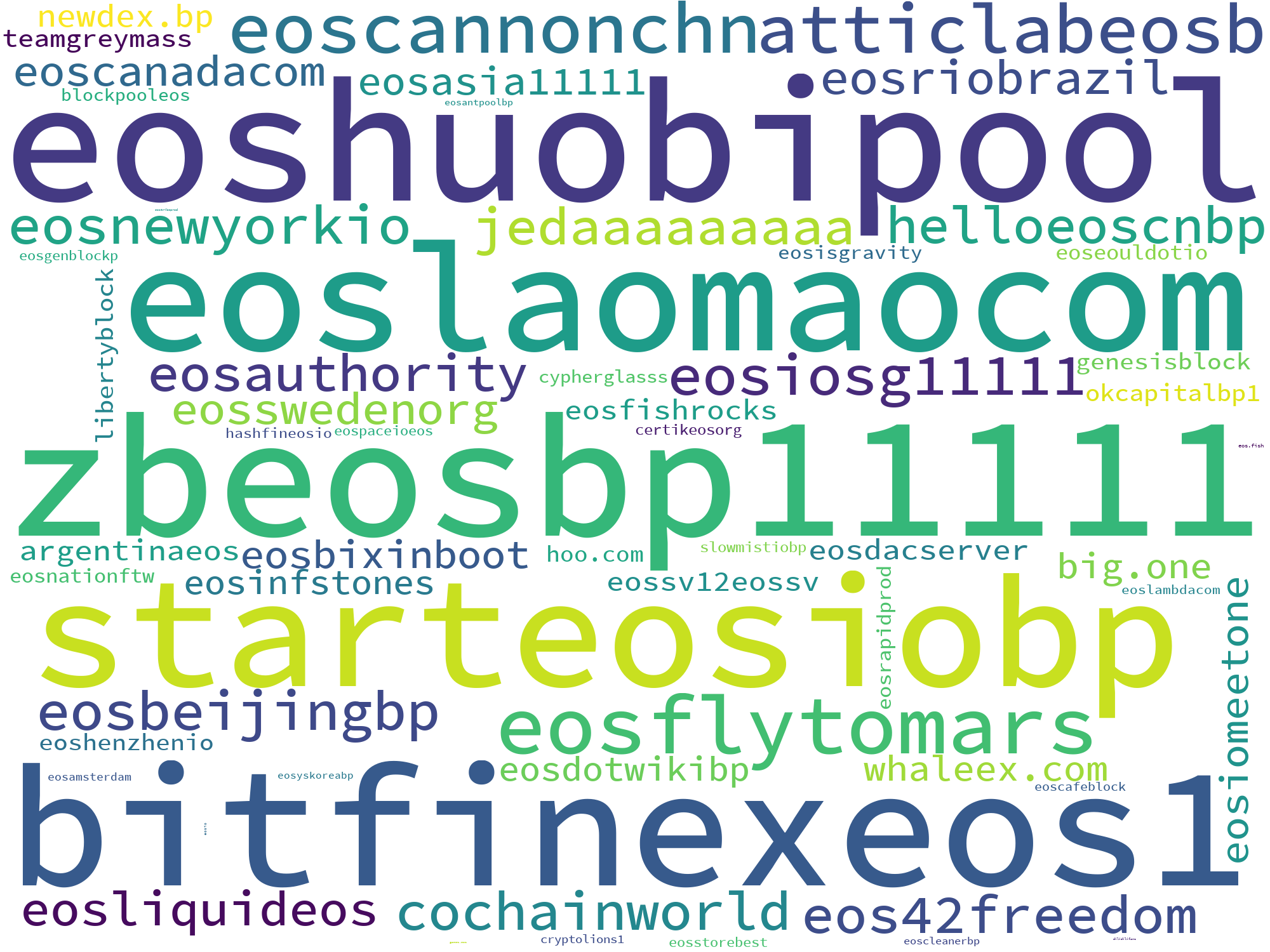}\label{1_BPInfo}
} 
\subfigure[\scriptsize Transaction and Action Counts]{
\centering
\includegraphics[width=5.3cm]{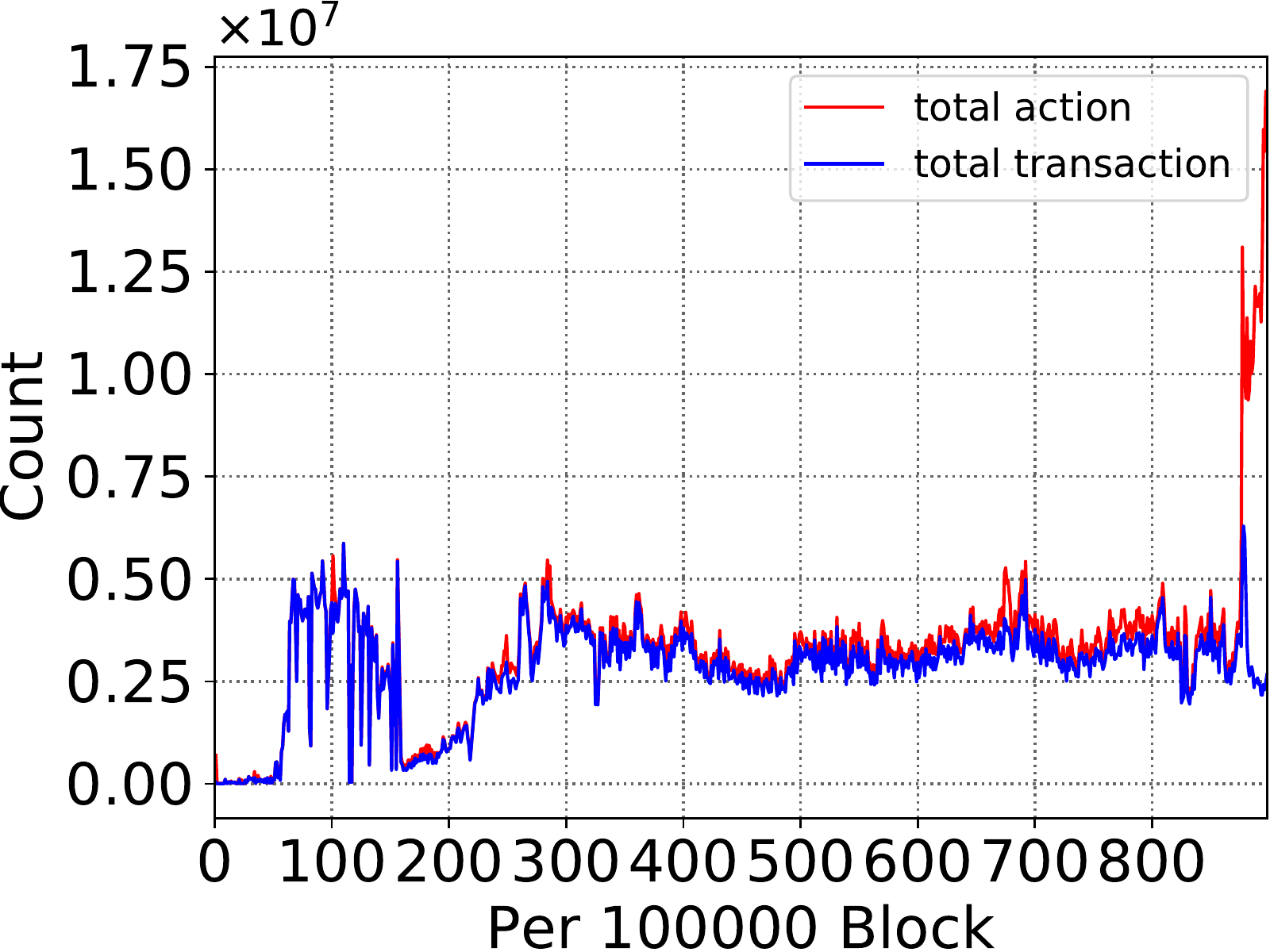}\label{1_txandac}
}
\subfigure[\scriptsize Deferred Transaction Count]{
\centering
\includegraphics[width=5.3cm]{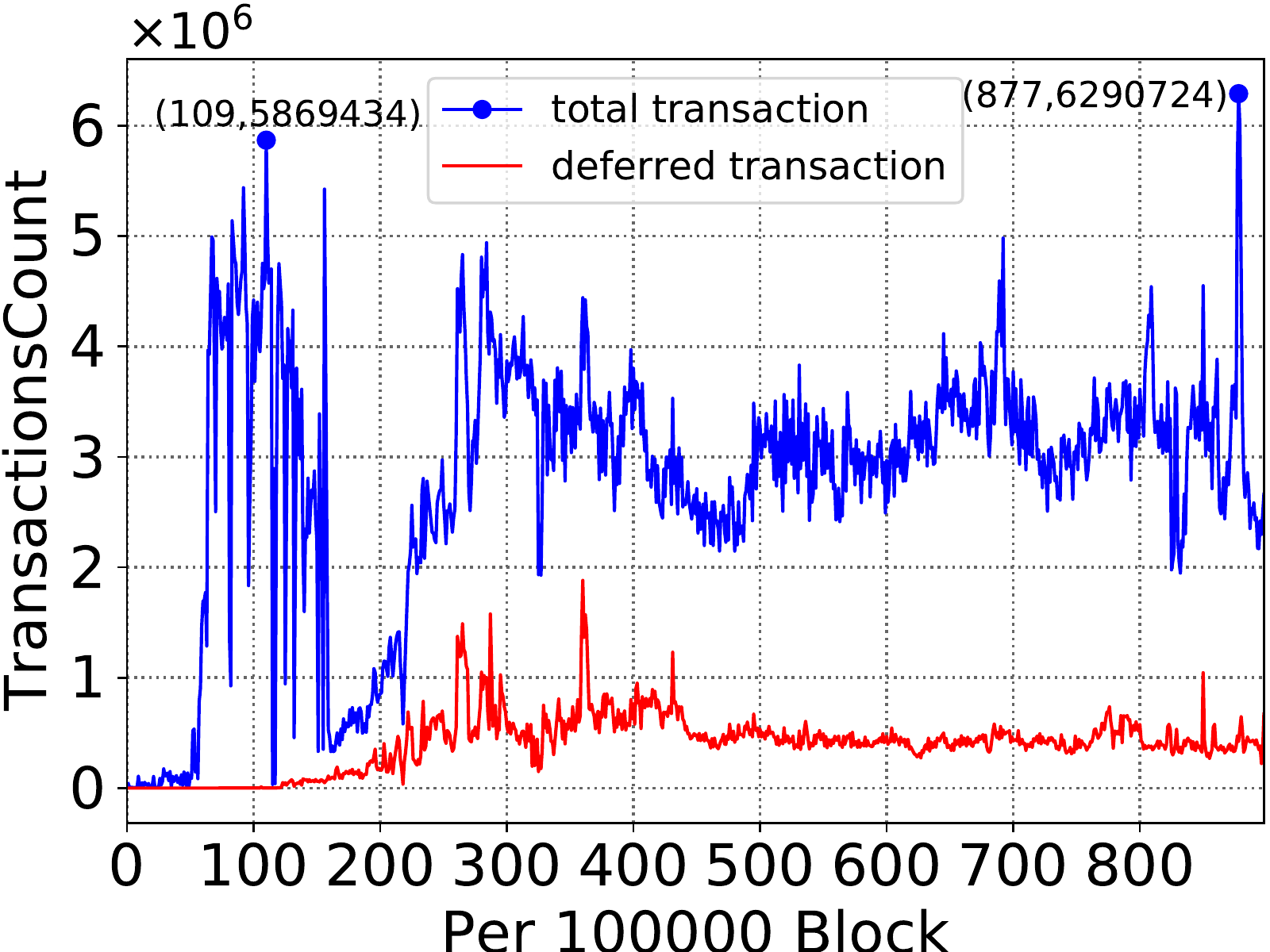}\label{1_txanddef}
} 
\subfigure[\scriptsize Usage of CPU and NET] {
\centering
\includegraphics[width=5.3cm]{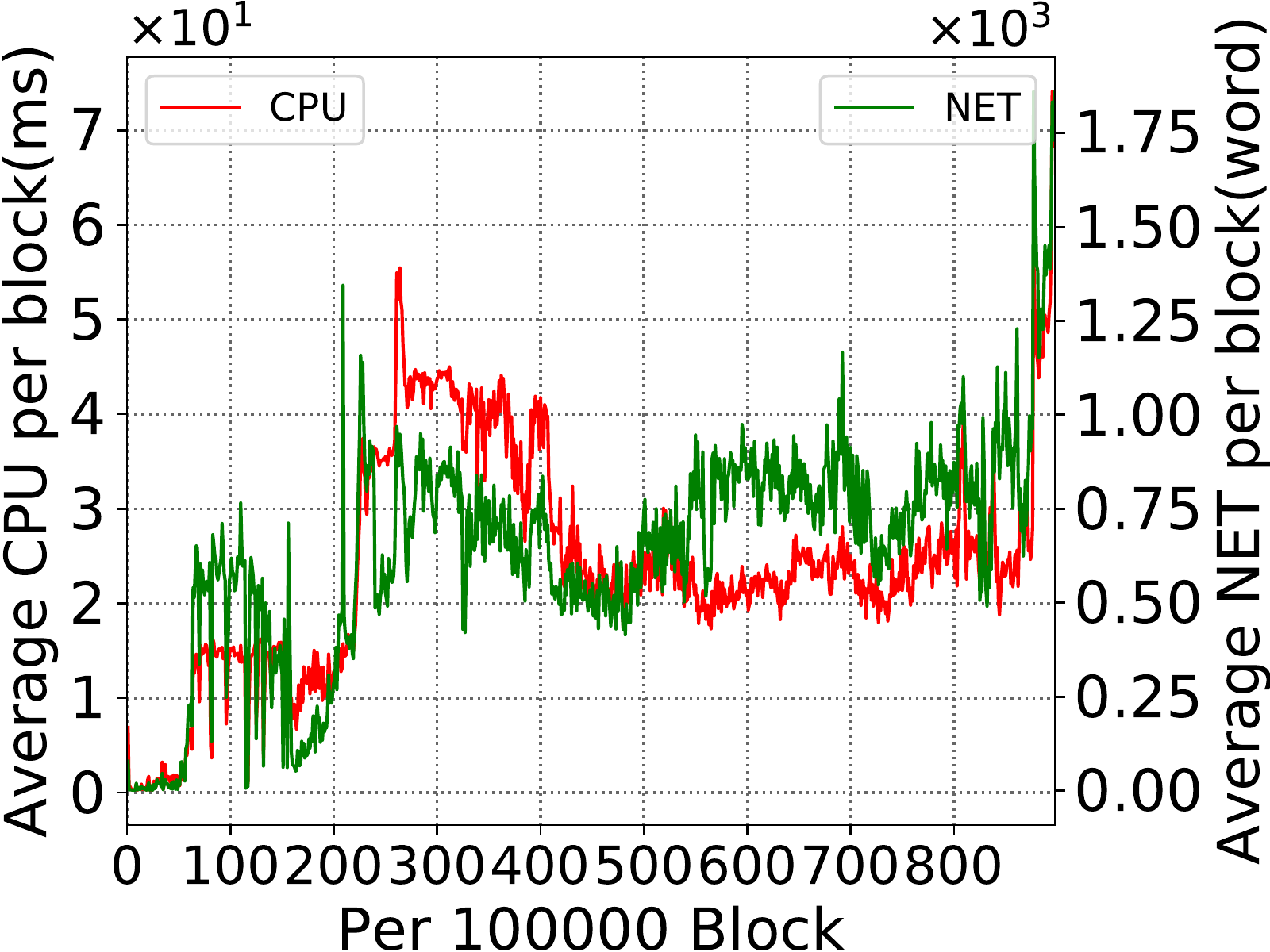}\label{1_cpuandnet}
}
\caption{Statistics of Dataset 1 (better viewed in color)} 
\vspace*{-0.25cm}
\end{figure}

There are only 63 unique block producers who however have generated 89,800,000 blocks, as shown in Table~\ref{Statistics of Dataset 1}. In contrast to Ethereum, there are 5,122 unique miner addresses generating 8,100,000 blocks~\cite{zheng2019xblock}. It implies that EOSIO does not have strong decentralization as Ethereum since few producers in EOSIO generate most of the blocks. Figure~\ref{1_BPInfo} shows the word-cloud statistics of account names of block producers in EOSIO. The word-cloud result also shows that several accounts almost dominate block production. These accounts have essentially been controlled by some exchanges, such as \texttt{eoshuobipool}, \texttt{zbeosbp11111}, and \texttt{bitfinexeos1}.
% These accounts have essentially been controlled by some exchanges, such as \emph{eosuobipool}\footnote{\url{https://www.huobi.io/en-us/}}, \emph{zbeosbp11111}\footnote{\url{https://www.zb.com/en/}}, \emph{bitfinexeos1}\footnote{\url{https://www.bitfinex.com/}}.

% 89800000
% 2533292528
% 357455192
% 2916530553

\renewcommand\subfigcapskip{-0.5ex}
As shown in Table~\ref{Statistics of Dataset 1}, the average number of transactions per block is 28.21. In other words, the average transaction throughput of EOSIO is 56.42 tps (transactions per second) since EOSIO produces a block every 0.5 seconds (i.e., $\mathrm{tps}=28.21/0.5=56.42$). When the network is active, as shown in Figure~\ref{1_txanddef} (when blocks reaching 87,700,000), the throughput can reach about 126 tps ($\approx\frac{6,290,724}{10^5\times0.5}$). It shows that EOSIO does achieve significant performance improvement compared with Bitcoin and Ethereum. However, EOSIO still has a long way to go to reach its goal of million-level tps.

In EOSIO, an operation that a user interacts with the blockchain is represented as an action, and a transaction can contain one or multiple actions. Figure~\ref{1_txandac} plots the count of actions (represented by the red curve) and the count of transactions (represented by the blue curve). It is shown in Figure~\ref{1_txandac} that, most of the time, the count of actions is quite close to that of transactions (i.e., the count of actions is only slightly higher than that of transactions), implying that every transaction nearly contains only one action. In addition, when blocks reaching 87,600,000, the count of actions surges; this effect was essentially caused by an EIDOS \emph{airdrop}. Anyone can transfer any amount of EOS to the contract \texttt{\small eidosonecoin}, and then receive the same amount of EOS and some EIDOS tokens from \texttt{\small eidosonecoin}. The amount of the obtained EIDOS tokens depends on the number of EOS-transfer actions but not on the transfer amount. Therefore, to gain more EIDOS tokens, many users include a number of transfer actions of 0.0001 EOS to \texttt{\small eidosonecoin} into a transaction.

EOSIO introduces a delayed communication mode to support initiating a transaction to be executed in the future. As shown in Table~\ref{Statistics of Dataset 1}, there are 357,455,192 deferred transactions, accounting for about 1/7 of the total transactions. Meanwhile, as shown in Figure~\ref{1_txanddef}, when blocks reaching 35,900,000, the count of deferred transactions is close to 1/2 of that of total transactions. This shows that deferred transactions are also a daily requirement for users to interact with the EOSIO blockchain. For example, the system contract \texttt{\small eosio.system} in EOSIO will trigger a delayed transaction \textit{refund} after the users redeem their mortgaged CPU or NET resources. The \textit{refund} transaction will return the corresponding EOS to the users after three days.

CPU and NET are necessary resources for transaction execution in EOSIO. CPU is used for computation, and NET is used for network transmission between block producers. Figure~\ref{1_cpuandnet} shows the statistics of average block CPU usage (in milliseconds) and NET usage (in words, and 1 word = 8 bytes) versus block count. We observe that both CPU usage and NET usage have similar trend to that of the count of transactions (as shown in Figure~\ref{1_txandac}). It is because the transaction count can directly affect both CPU and NET usage. However, we can see that the changing amplitude of CPU usage is different from that of NET usage at some moments. For example, the increment of CPU usage is higher than that of NET usage when blocks falling into the range from 23,0000,00 to 43,000,000. This is due to the increment in deferred transactions as shown in Figure~\ref{1_txanddef}. Generally, a deferred transaction is triggered by another original transaction. When the block producers receive the original transaction, they can generate the content of the deferred transaction during the execution process and saved them locally for future execution. Therefore, deferred transactions require almost no NET resources, whose contents rarely need to be transmitted over the network.

\subsection{Dataset 2: Internal and External EOS Transfer Action} \label{Dataset 2: Internal and External EOS Transfer Action}

%图 以太币转账金额分布图√,  以太币转账金额折线图√
EOS, as the most representative cryptocurrency of EOSIO, was created and issued by the contract account \texttt{\small eosio.token} when the EOSIO \textit{mainnet} was launched. In EOSIO, EOS transfers can be divided into \emph{external} transfers and \emph{internal} transfers. In general, an external transfer action represents a direct transfer from users to users (or from users to contracts), which is recorded in the transaction of the block. The internal transfer action is essentially an inline action, which is triggered by another action and is not be observed in the block. For example, when a user buys RAM or stakes EOS for CPU and NET, an inline action that transfers EOS to the system account (i.e., \texttt{\small eosio.ram} and \texttt{\small eosio.ramfee}) is triggered. As shown in Table~\ref{Statistics of Dataset 2}, there are 1,356,748,049 internal transfers and 653,529,552 external transfers that occur among 1,156,658 accounts. 

% 653529552 + 1356748049 + 99999990.01 + 1156658 + 19380114743.780087 / 2010277601

\renewcommand\subfigcapskip{-0.5ex}
\begin{figure}
\centering
\subfigure[\scriptsize EOS External Transfer Amount]{
\centering
\includegraphics[width=5.3cm]{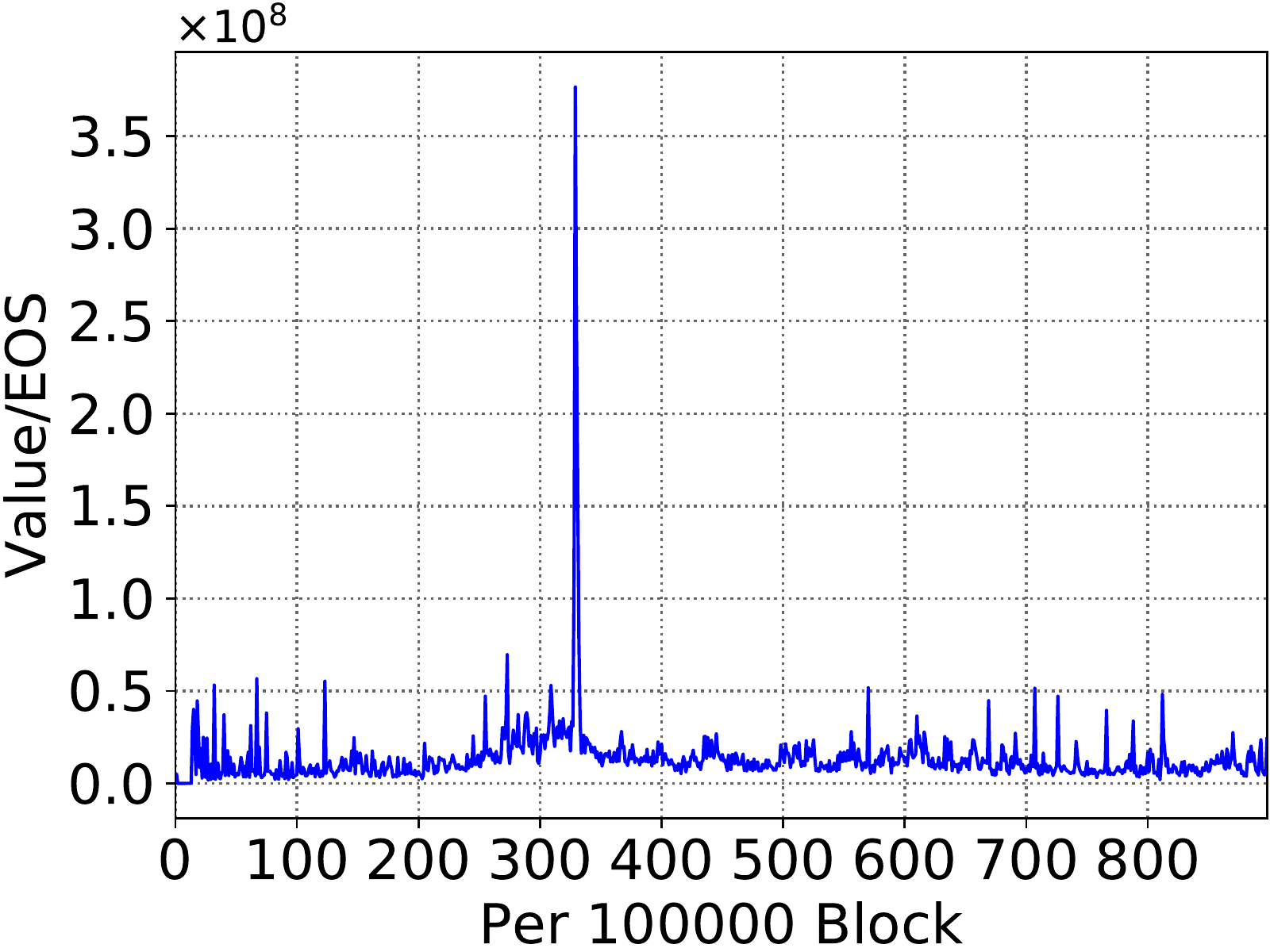} \label{2_external}
}
\subfigure[\scriptsize EOS Internal Transfer Amount]{
\centering
\includegraphics[width=5.03cm]{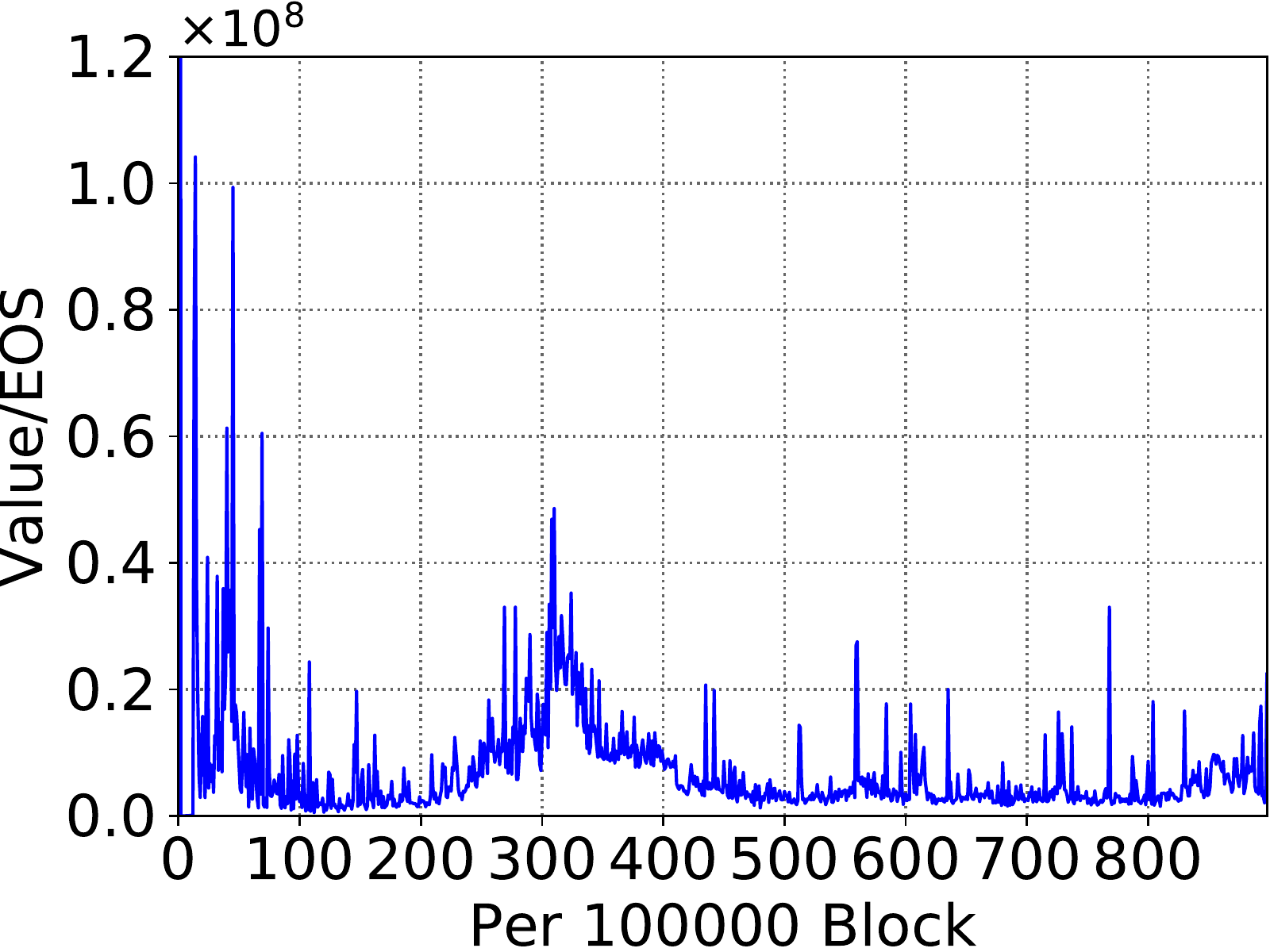} \label{2_internal}
}
\subfigure[\scriptsize EOS Transfer Amount Distribution]{
\centering
\includegraphics[width=5.3cm]{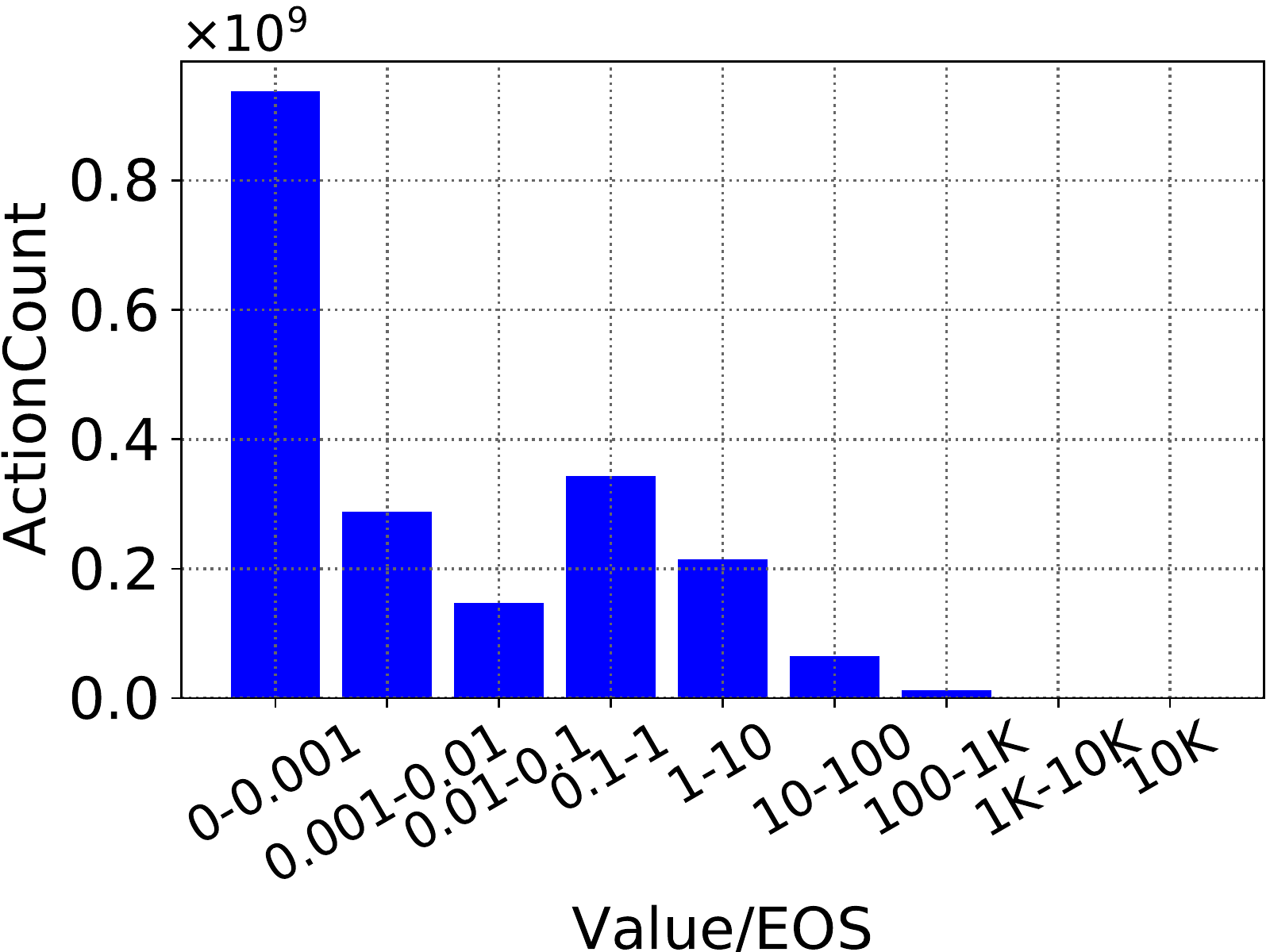} \label{2_eostrandis}
}
\subfigure[\scriptsize Word Cloud Statistics of Transfer Memo]{
\centering
\includegraphics[width=5.3cm]{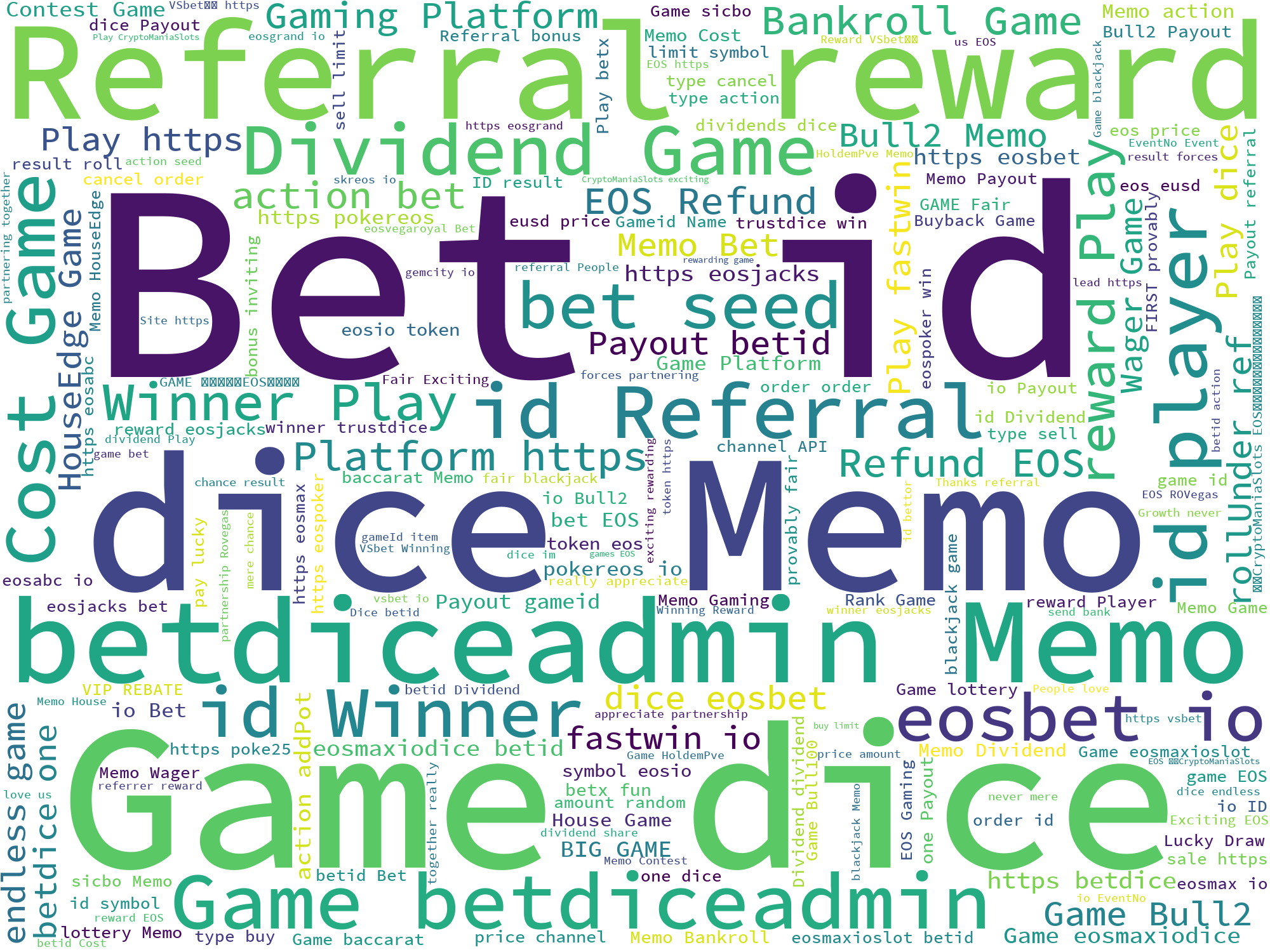} \label{2_eosmemo}
\vspace*{-0.25cm}
}

\caption{Statistics of Dataset 2 (better viewed in color)}
\end{figure}
Figures~\ref{2_external} and Figure~\ref{2_internal} show the total internal transaction amount and the total external transaction amount of every 100,000 blocks, respectively. It can be seen that EOS internal and external transfers are active around the block 33,000,000, matching with the most active time of the Gambling and Games DApps~\cite{min2019blockchain}. In EOSIO, users have the right to write memos into an EOS transfer action. Figure~\ref{2_eostrandis} shows the visualization of the word-cloud statistics of EOS transfer memos. The result shows that many memos contain words related to gambling and games, such as \textit{Bet}, \textit{Dice}, \textit{Game}, etc. It implies that the Gambling and Games DApps are prevalent in EOSIO, which is attributable to free charge for transferring money in EOSIO.

\begin{table}[h]
    \setlength{\belowcaptionskip}{0.2cm}
    \centering
    \caption{Statistics of Dataset 2}
    \renewcommand{\arraystretch}{1.5}
    \footnotesize
    \begin{tabular}{|l|r|}
    \hline
    \textbf{Statistics} &  \textbf{Values} \\ \hline\hline
    No. of Internal EOS Transfers     &   1,356,748,049 \\ \hline    %???
    No. of External EOS Transfers     &  653,529,552 \\ \hline    %???
    No. of Accounts     &  1,156,658 \\ \hline               %???
    Mean Amount of EOS     &  9.64 \\ \hline
    Maximum Amount of EOS     &  99,999,990.01\\ \hline
    \end{tabular}
    \label{Statistics of Dataset 2}
\vspace*{-0.25cm}
\end{table}

The values of EOS have a large variance as the maximum value is 99,999,990.01 EOS (about 100 million dollars when writing this paper), but the mean is only 9.64 EOS as shown in Table~\ref{Statistics of Dataset 2}. The distribution of EOS transfer amount is shown in Figure~\ref{2_eostrandis}. We can find that most EOS transfer actions fall into the range from 0.0001 to 10, and very few transfer actions exceed 1,000 EOS, indicating that most transfer actions in EOSIO only transfer a small amount of EOS.

\subsection{Dataset 3: Contract Information} \label{Dataset 3: Contract Information}
Similar to Ethereum, EOSIO also supports Turing-complete smart contracts. Users can deploy a new contract on an account through the interface \texttt{\small SetCode} of the system account \texttt{\small eosio}. It is worth noting that users can easily update or delete contract code with the same interface while this action is not allowed in Ethereum. In order to investigate all smart contracts in EOSIO, we process the raw data to obtain basic information about the smart contracts, including the \textit{creation (update) time}, \textit{contract code}, and \textit{code size}. Here, we name the action of setting contracts code to empty through \texttt{\small SetCode} as \texttt{\small SetEmptyCode}, being equivalent to the removal of the contract (the contract can be deployed again on the same account later).

\begin{figure}[t]
\centering 
\subfigure[Contract code size distribution]{
\centering
\includegraphics[width=5.3cm]{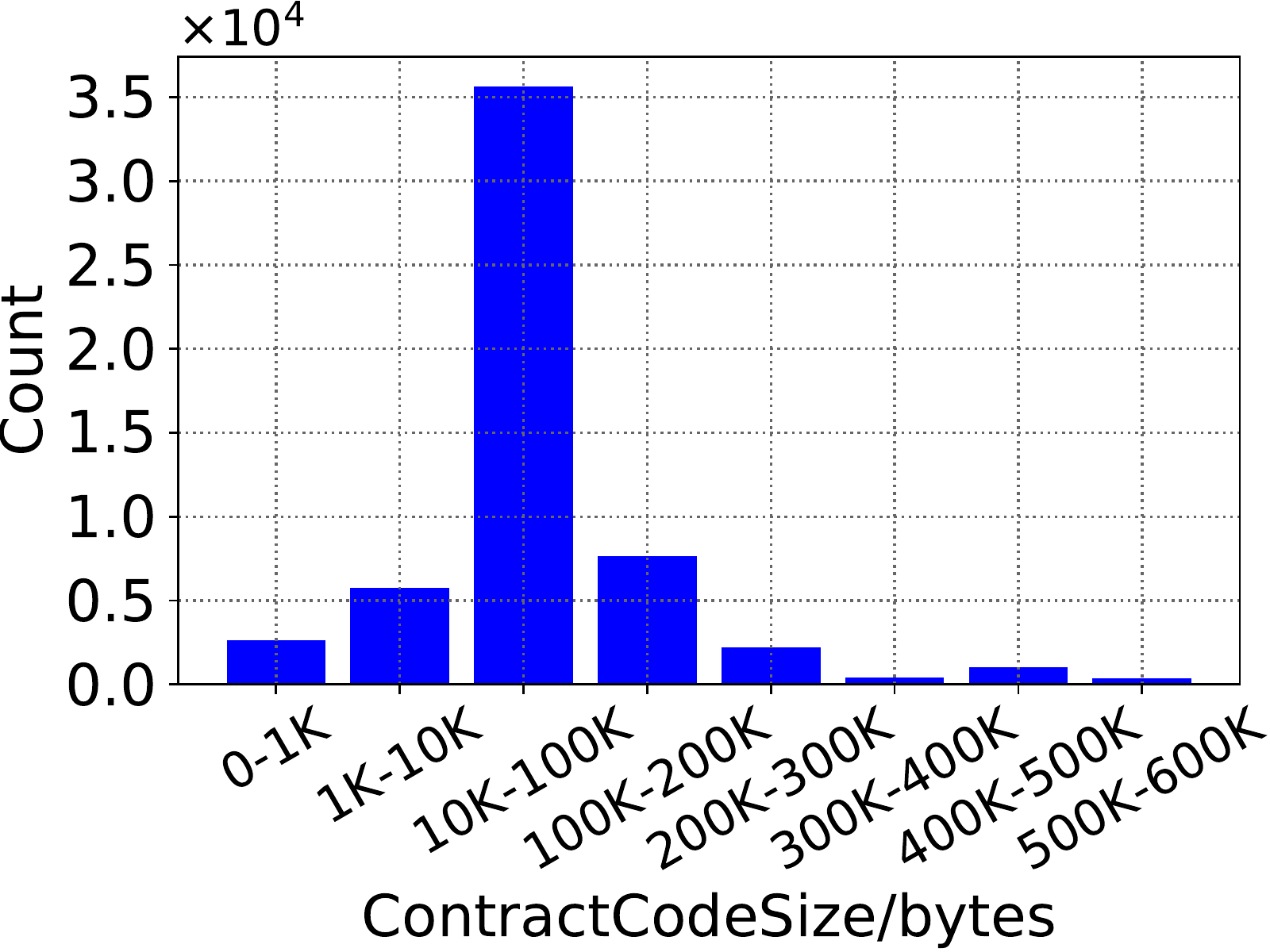}\label{3_contractsize}
}
\subfigure[Count of created contracts ]{
\centering
\includegraphics[width=5.3cm]{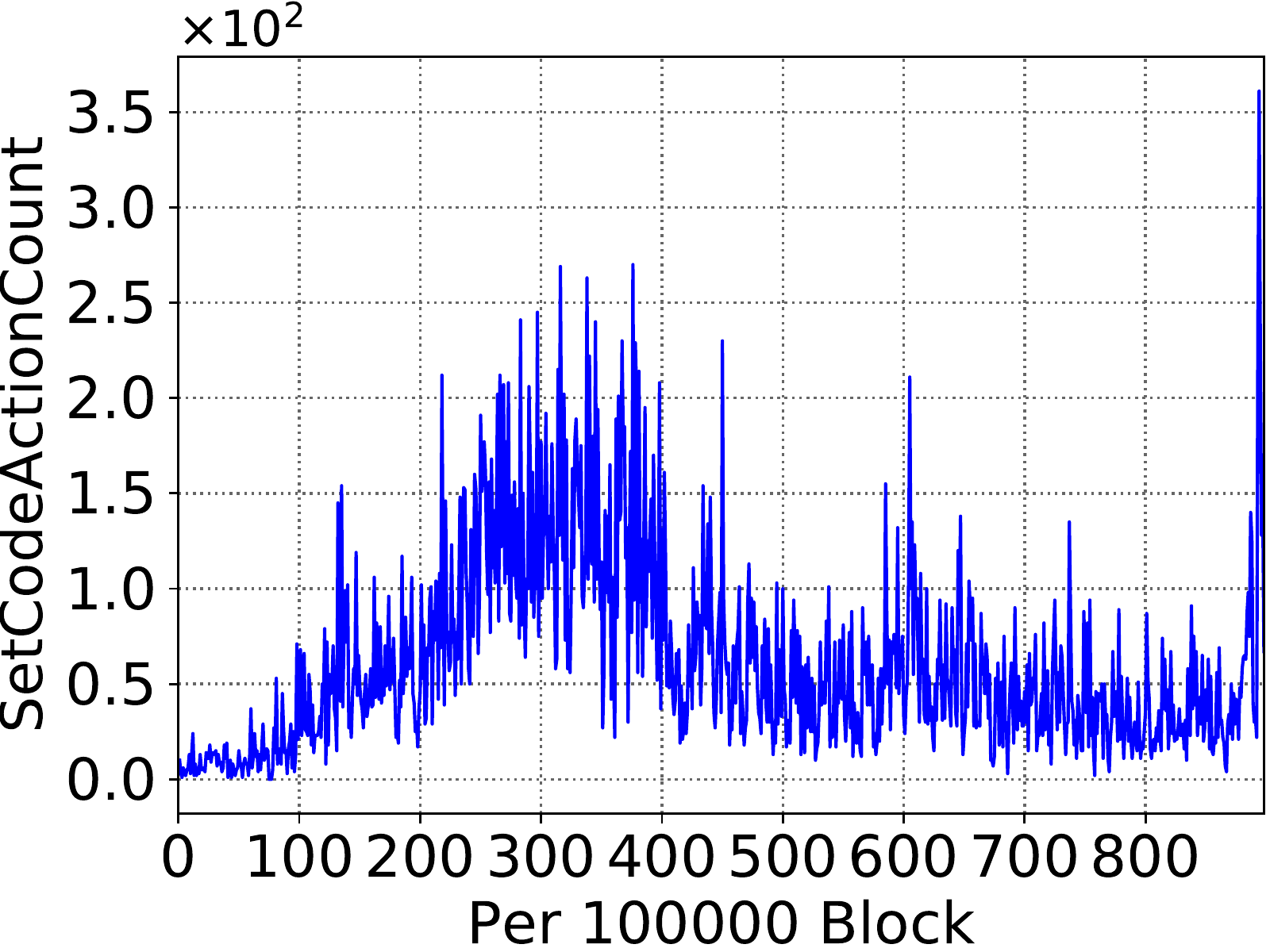}\label{3_contracttime}
}
\caption{Statistics of Dataset 3}
\vspace*{-0.25cm}
\end{figure}

According to the statistics as shown in Table~\ref{Statistics of Dataset 3}, there are only 5,594 contracts, but there are 55,735 \texttt{\small SetCode} actions and 1,747 \texttt{\small SetEmptyCode} actions. It means that most contracts have been updated multiple times after deploying. In addition, the number of contracts in EOSIO is much smaller than that in Ethereum~\cite{zheng2019xblock}, because deploying a contract in EOSIO needs to buy expensive RAM to store the contract code. Figure~\ref{3_contracttime} shows the statistics of the total number of \texttt{\small SetCode} actions of every 100,000 blocks. It can be seen that when blocks reaching around 33,000,000, the creation or update of contracts is most active; this phenomenon matches the time when the gambling and game project parties launched a large number of new games or activities~\cite{min2019blockchain}.

\begin{table}[h]
    \setlength{\belowcaptionskip}{0.2cm}
    \centering
    \caption{Statistics of Dataset 3} \label{Statistics of Dataset 3}
    \renewcommand{\arraystretch}{1.5}
    \footnotesize
    \begin{tabular}{|l|r|}
    \hline
    \textbf{Statistics} &  \textbf{Values} \\ \hline\hline
    No. of Created Contracts     &   5,594  \\ \hline    
    No. of Contract \texttt{\small SetCode} Actions     &   55,735\\ \hline  
    No. of Contract \texttt{\small SetEmptyCode} Actions   &  1,747   \\ \hline    
    Mean of Contract Hex Code Size     &  75,470.34  \\ \hline
    \end{tabular}
\vspace*{-0.25cm}
\end{table}

\renewcommand\subfigcapskip{-0.5ex}

Regarding the contract code, we convert the contract code into hexadecimal code and obtain the size. Figure~\ref{3_contractsize} shows the statistics of the contract code size distribution. In particular, the average contract code size is 75,470.34 bytes, which is much larger than that of Ethereum~\cite{zheng2019xblock}. It is because there are more simple test contracts (with small size) in Ethereum while fewer test contracts are deployed in EOSIO. In addition, we observe that most contracts fall into the size with a range from 10k bytes to 100k bytes, implying that many contracts may look similar to each other. We will show that these contracts are related to token and gambling in the exploration of Section~\ref{Dataset 4: Contract Invocation}, further confirming the fact that there are certain similarities between the contracts.

\subsection{Dataset 4: Contract Invocation} \label{Dataset 4: Contract Invocation}

% 2189162705+775082+14751
% 1183985192 2189162705
% 0.540839284944789

\begin{table}[h]
    \setlength{\belowcaptionskip}{0.2cm}
    \centering
    \caption{Statistics of Dataset 4}
    \renewcommand{\arraystretch}{1.5}
    \footnotesize
    \label{Statistics of Dataset 4}
    \begin{tabular}{|l|r|}
    \hline
    \textbf{Statistics} &  \textbf{Values} \\ \hline\hline
    No. of Contract Invocation Actions     &   2,189,162,705  \\ \hline   
    No. of Calls with Errors    &   14,751   \\ \hline 
    No. of Authorization accounts     &  775,082 \\ \hline  
    \end{tabular}
\end{table}

Unlike Ethereum, all actions (transactions) in EOSIO are completed through calling contracts, including common EOS transfers. There are several system contract accounts in EOSIO, such as \texttt{\small eosio}, \texttt{\small eosio.token}, \texttt{\small eosio.msig}, and so on. These system accounts are responsible for the daily affairs in EOSIO, such as transferring EOS, buying RAM, staking CPU or NET, etc. In order to investigate the contract ecology of EOSIO, we extract the invocation data of all contracts except the system contracts. The contract invocation dataset includes \textit{calling time}, \textit{authorizer}, \textit{called contract}, and \textit{calling function}. As shown in Table~\ref{Statistics of Dataset 4}, 775,082 authorization accounts initiated a total number of 2,189,162,705 contract invocations, among which 14,751 contract invocations contain errors.

\renewcommand\subfigcapskip{-0.5ex}
\begin{figure}[t]
\centering 
\subfigure[\scriptsize  Count of Contract Invocations]{
\centering
\includegraphics[width=5.3cm]{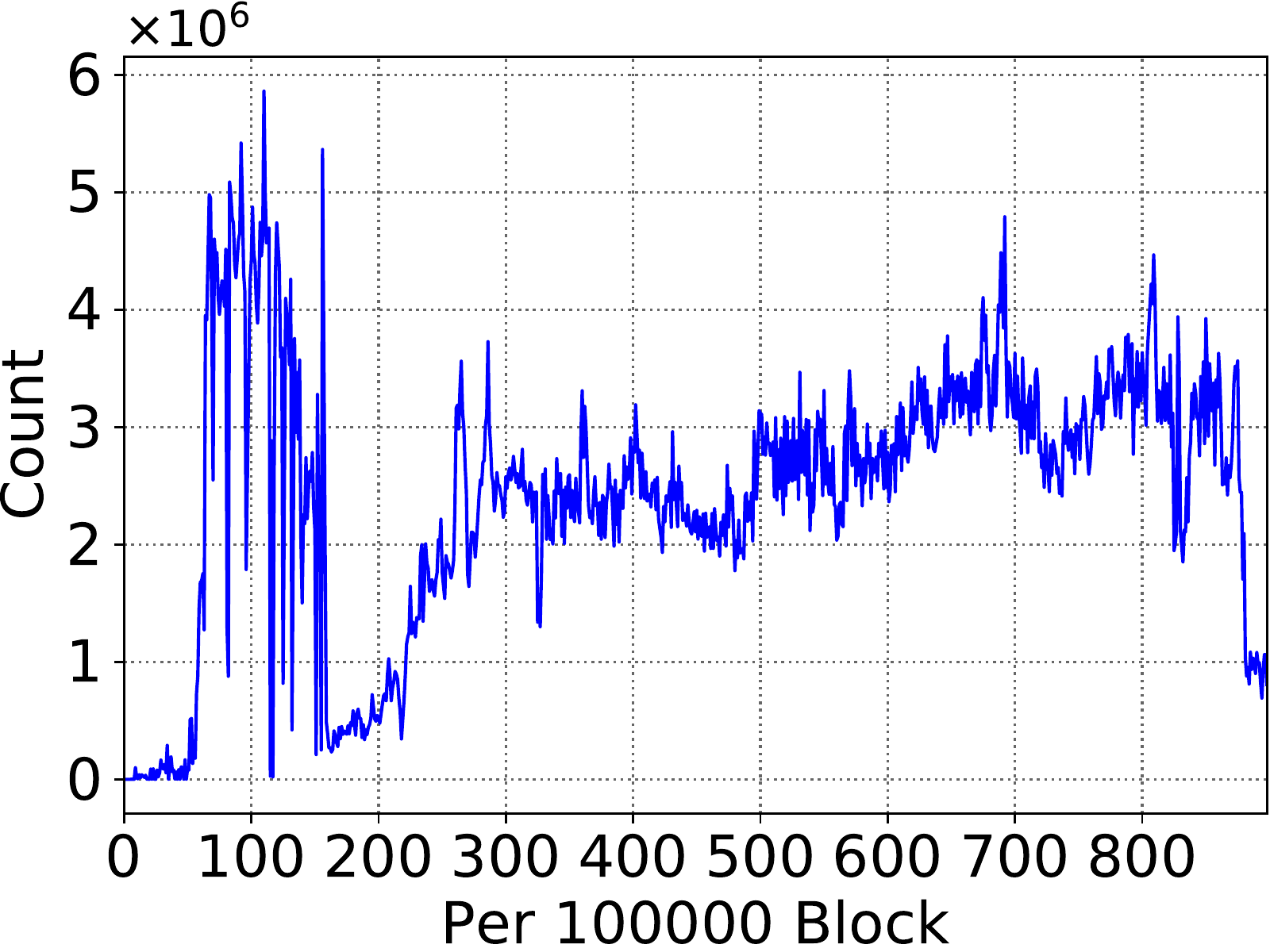}\label{4_calltime}
} 
\subfigure[\scriptsize Call Type Distribution]{
\centering
\includegraphics[width=5.3cm]{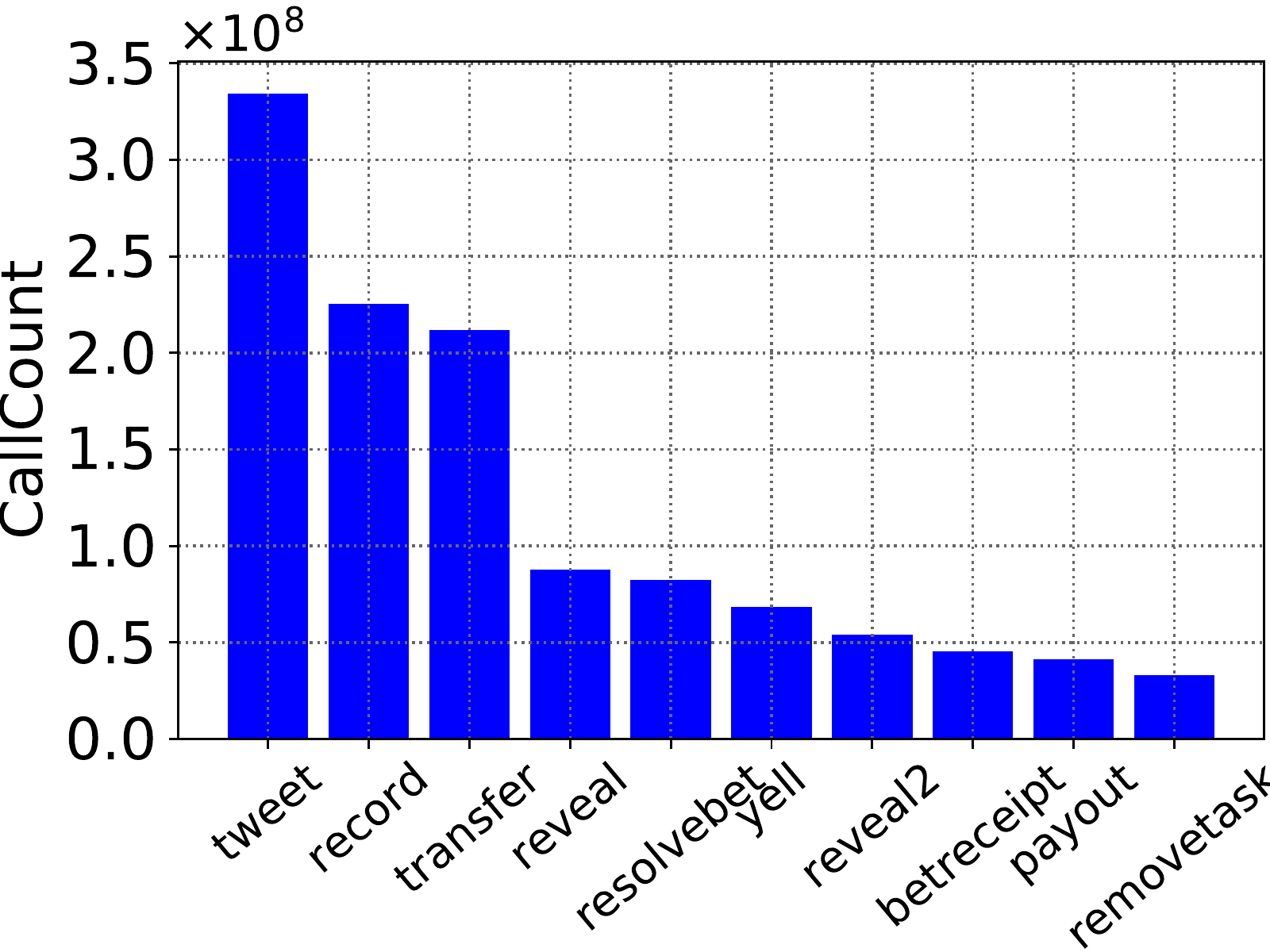}\label{4_calltype}
}
\caption{Satistics of Dataset 4} \label{Visualization of Dataset 4}
\end{figure}

Figure~\ref{4_calltime} shows the count of contract invocations of every 100,000 blocks. It can be seen that when blocks are in the interval from 5,000,000 to 12,000,000, the count of contract invocations has a periodic peak. It is because a contract namely \textit{blocktwitter} periodically launches a large number of actions named \textit{tweets} for pressure testing, only carrying a message ``\textit{WE LOVE BM}''. Figure~\ref{4_calltype} shows the top-10 most frequently-called functions, which account for 54.08\% of all contract invocations. It indicates that most of the calling functions concentrate on some of these frequently-called functions. In addition, the function \texttt{\small tweet} mentioned above ranks the first, while the function \texttt{\small transfer} related to tokens ranks the third.  We also found that the functions related to Gambling and Games DApps, such as \texttt{\small reveal}, \texttt{\small resolvebet}, and \texttt{\small reveal2}, also appear in the top-10 functions. Literally, these functions often represent lottery actions for Gambling and Games DApps.

\subsection{Dataset 5: Token Action} \label{Dataset 5: Token Action}
%图 Token活跃度分布图√ Token名字词云√
\begin{table}[h]
    \setlength{\belowcaptionskip}{0.2cm}
    \centering
    \caption{Statistics of Dataset 5}
    \renewcommand{\arraystretch}{1.5}
    \footnotesize
    \begin{tabular}{|l|r|}
    \hline
    \textbf{Statistics} &  \textbf{Values} \\ \hline\hline
    No. of Token Contracts     &      1,826 \\ \hline
    No. of Created and Issued Tokens     &  4,811  \\ \hline    
    No. of Token Transfer Actions     &  1,128,111,142   \\ \hline  
    No. of Holder Accounts     &  1,295,389\\ \hline  
    \end{tabular}
    \label{Statistics of Dataset 5}
\end{table}

From the prior analysis in Section~\ref{Dataset 3: Contract Information} and Section~\ref{Dataset 4: Contract Invocation}, we observe that token contracts are active in EOSIO. Next, we will further investigate token contracts. In EOSIO, a contract that contains three standard functions: \texttt{\small create}, \texttt{\small issue}, and \texttt{\small transfer} can be regarded as a standard token contract. According to this condition, we extract the token action dataset from the raw data. The token action dataset contains basic information for each token, including \textit{name (symbol)}, \textit{creation time}, \textit{issuer}, \textit{total issued amount}, and so on.

As shown in Table~\ref{Statistics of Dataset 5}, 1,826 contracts are considered as standard token contracts, and a total of 4,811 tokens have been created and issued. It implies that in EOSIO, a contract can issue multiple tokens, which is different from that of Ethereum. In addition, a total of 1,128,111,142 token transfers occurred in 1,295,389 holding accounts. Generally, the number of holding accounts is larger than the exact number of human holders, because a real-world user often has several accounts. In addition, token issuers can send tokens directly to any account without permission, being commonly known as \textit{Token Airdrop}.

\renewcommand\subfigcapskip{-0.5ex}
\begin{figure}[t]
\centering 
\subfigure[\scriptsize  ERC20 Popularity Distribution]{
\centering
\includegraphics[width=5.3cm]{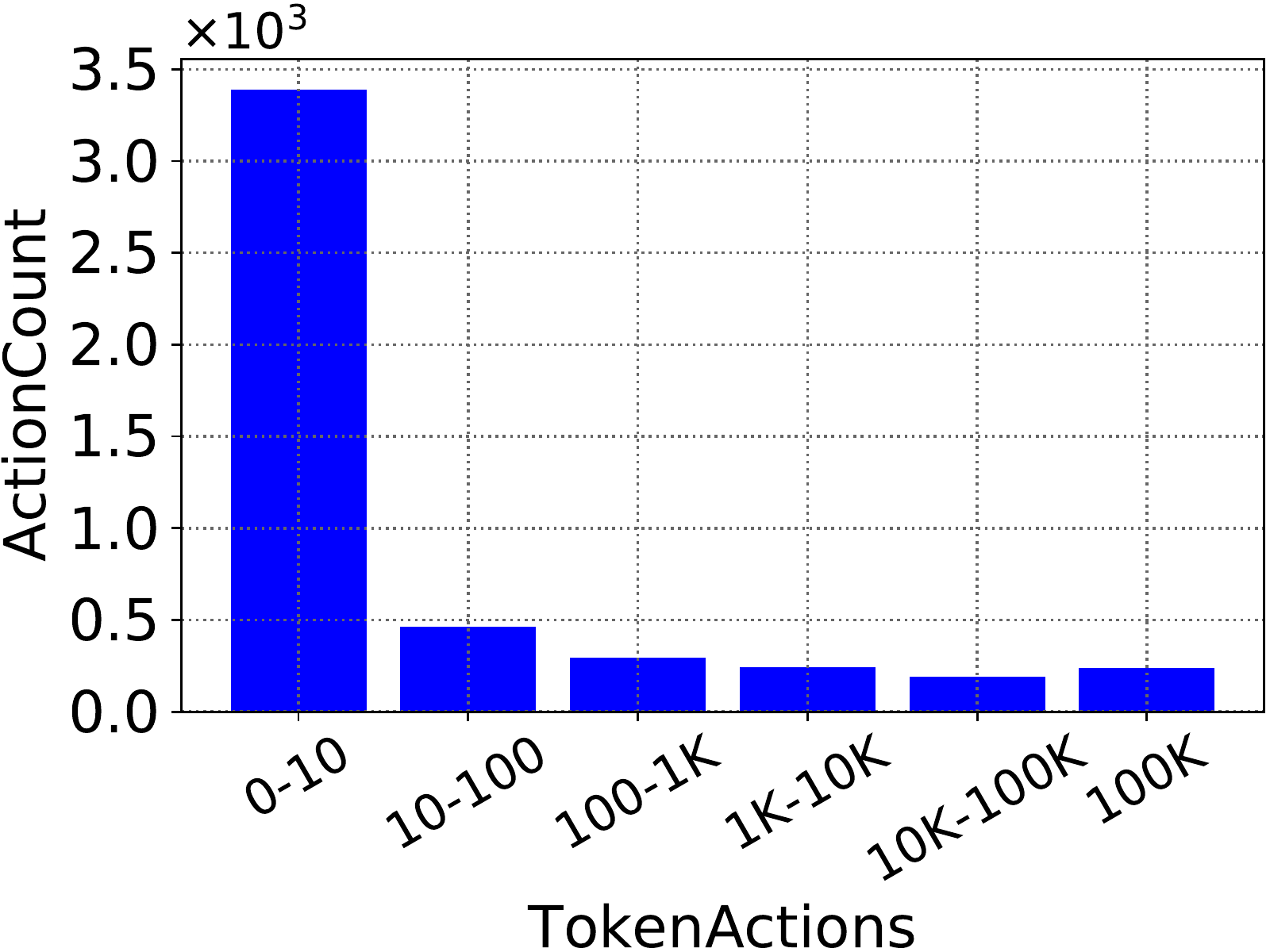} \label{5_erc20count}
}
\subfigure[\scriptsize Word-Cloud Statistics of ERC20 Tokens]{
\centering
\includegraphics[width=5.3cm]{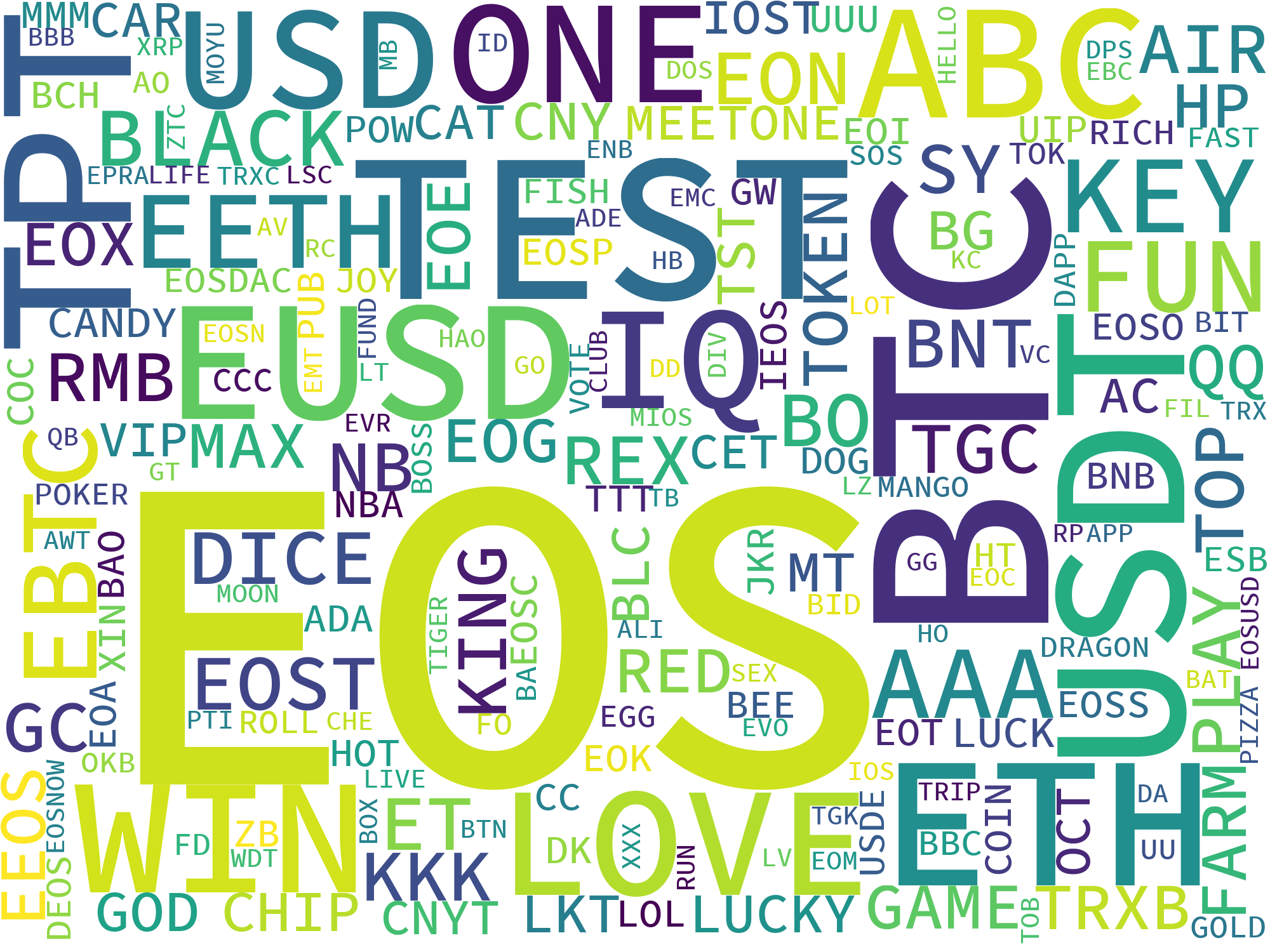} \label{5_erc20Cloud}
} 
\caption{Statistics of Dataset 5 (better viewed in color)}
\end{figure}

Figure~\ref{5_erc20count} shows the distribution of the transfer count of each standard type of token in EOSIO. We can easily observe the Matthew effect \cite{merton1968matthew} from Figure~\ref{5_erc20count}: the rich get richer, as most of the transfer actions occur on a few token contracts. Up to 80.02\% of token contracts have less than 100 transfers. Figure~\ref{5_erc20Cloud} shows the word-cloud statistics of token names. It can be seen that the most common word is \textit{EOS}, which is the name of the native cryptocurrency of EOSIO. Other common words are \textit{BTC}, \textit{ETH}, \textit{USDT}, etc, which are the names of well-known cryptocurrency tokens. Meanwhile, we have also found that many token names contain the word \textit{TEST}, indicating that these token contracts are used for testing.

\subsection{Dataset 6: Account Creation}\label{Dataset 6: Account Creation}
\begin{table}[h]
    \setlength{\belowcaptionskip}{0.2cm}
    \centering
    \caption{Statistics of Dataset 6}
    \renewcommand{\arraystretch}{1.5}
    \footnotesize
    \begin{tabular}{|l|r|}
    \hline
    \textbf{Statistics} &  \textbf{Values} \\ \hline\hline
    No. of \textit{NewAccount} Actions    & 1,636,043   \\ \hline    
    No. of Creator    & 45,350   \\ \hline    
    \end{tabular}
    \label{Statistics of Dataset 6}
\end{table}

In most public blockchain systems, creating a new address (account) is easy and free. However, in EOSIO, creating a new account requires a creator to buy RAM for storing account information. In addition, the creator will generally stake some CPU and NET resources for the new account to initiate transactions. In order to investigate the account creation in EOSIO, we extract the account creation dataset from the raw data, which mainly includes the \textit{creation time}, \textit{creator}, \textit{account name}, etc. It is worth noting that, in EOSIO, the account names are allowed with up to the 12-character length and each character is only allowed to use either digits \texttt{[1-5]} or alphabet letters \texttt{[a-z]}. As shown in Table~\ref{Statistics of Dataset 6}, there are  1,636,043 different accounts (\textit{or NewAccount}s), which were created by only 45,350 account creators. It shows that one account creator can create a large number of accounts. Specifically, each creator in EOSIO created 36.08 accounts on average. In addition, the number of the created accounts is much larger than that of the authorization accounts or token holders mentioned in the above analysis, also implying that many accounts are not active in EOSIO. 

\renewcommand\subfigcapskip{-0.5ex}
\begin{figure}[t]
\centering 
\subfigure[\scriptsize Created Account Count]{
\centering
\includegraphics[width=5.3cm]{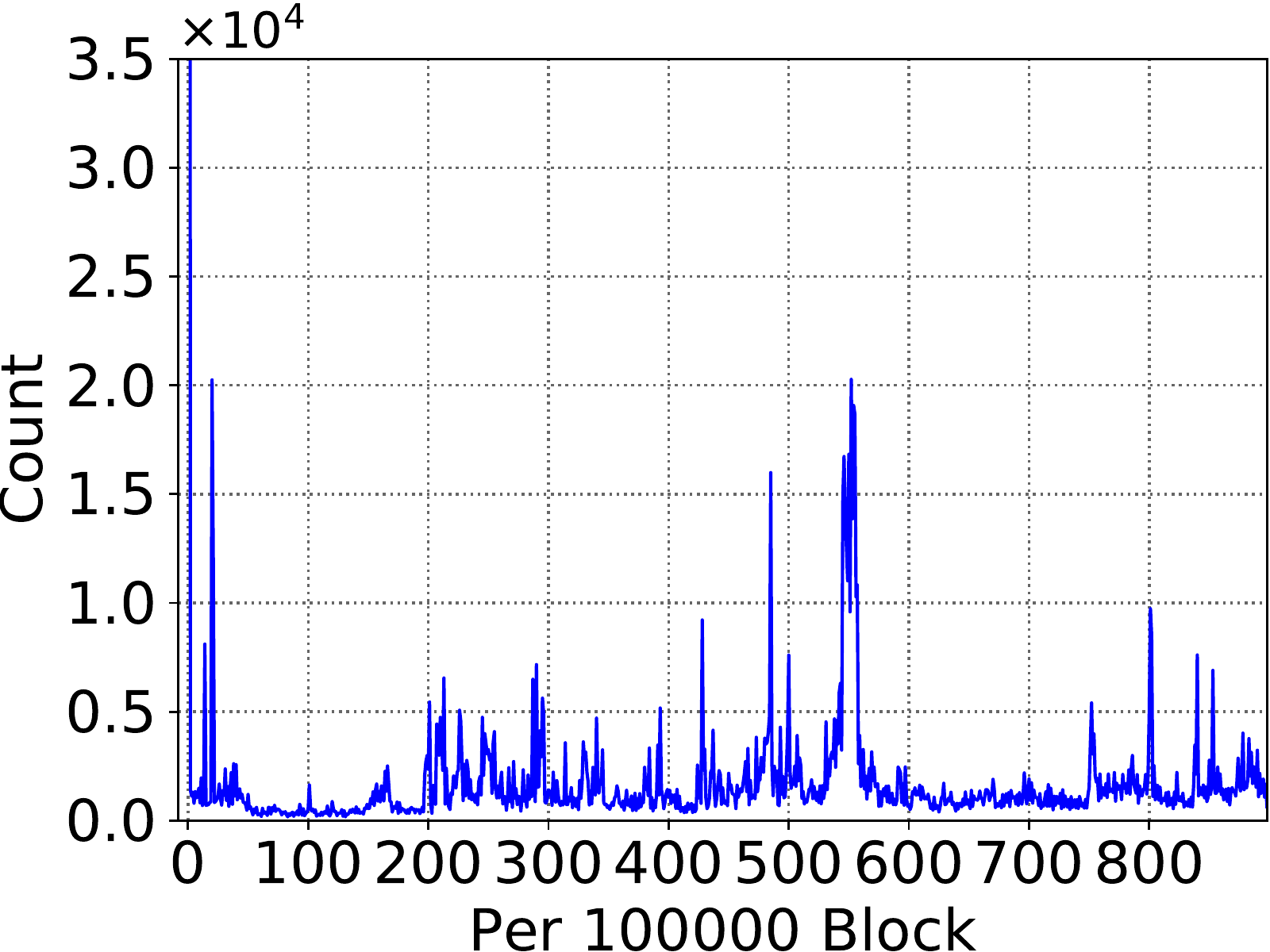} \label{6_newacc}
}
\subfigure[\scriptsize Work Cloud Statistics of Account Names]{
\centering
\includegraphics[width=5.3cm]{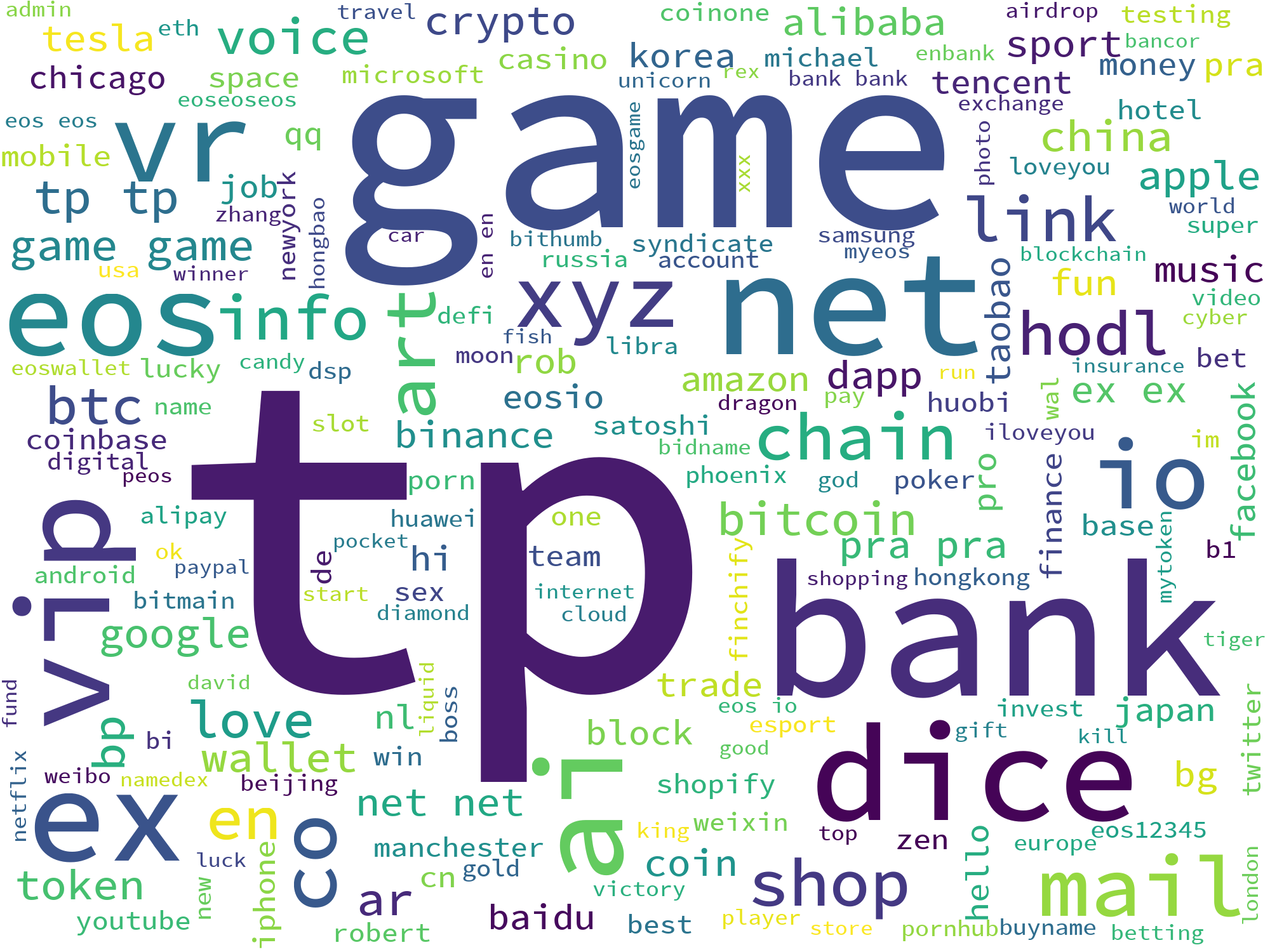} \label{6_accName}
} 
\caption{Statistics of Dataset 6 (better viewed in color)}
\end{figure}

Figure~\ref{6_newacc} shows the created accounts of every 100,000 blocks. It can be seen that a large number of accounts were created during the initial launch of the EOSIO \textit{mainnet}. These accounts are almost created by the system account \texttt{\small eosio} and their names are similar to each other, implying that they are mainly used for testing. For example, many accounts are prefixed with \texttt{\small hex} or \texttt{\small ha2}. For most of the time, the created accounts of every 100,000 blocks fall roughly in the range from 1,000 to 5,000, except for the time around the block 55,000,000. This time coincided with the time when EOSIO's total account number exceeded one million and EOSIO officially launched the REX mechanism. Figure~\ref{6_accName} shows the word-cloud statistics of account names. We can find that most accounts name contain \textit{tp} because a company called TokenPocke (TP) provides wallet services such as account creation in EOSIO. In addition, many accounts contain words, such as \textit{game}, \textit{dice}, \textit{bank}, etc., further implying that the Gambling and Games DApps are quite popular in EOSIO community.

%图 Token活跃度分布图√ 加密猫换手次数分布图√

\subsection{Dataset 7: Resource Management}\label{Dataset 7: Resource Management}
Unlike most public blockchain systems (such as Ethereum and its variants) that adopt \emph{gas} mechanism, EOSIO prevents malicious behaviors of contracts by limiting RAM, CPU, and NET resources. Users need to buy RAM to store information in EOSIO. The price of RAM is mainly determined by the supply-and-demand model of the market, and its core is the \emph{bancor protocol}~\cite{hertzog2017bancor}. In addition, users need to stake CPU and NET for transaction calculation and network transmission. The amount of CPU or NET resources obtained by users is mainly determined by the proportion of EOS staked by them to that of the entire network. Since the EOSIO \textit{mainnet} went live, the problem of insufficient CPU to complete even the simplest transfers has been criticized. Therefore, EOSIO officially launched the REX mechanism on May 1, 2019 to support the leasing service of CPU and NET to alleviate the problem. Users who cannot stake sufficient CPU or NET resources can rent from others in the system.

\begin{figure}
\centering 
\subfigure[\scriptsize Staked CPU and NET Amount]{
\centering
\includegraphics[width=5.0cm]{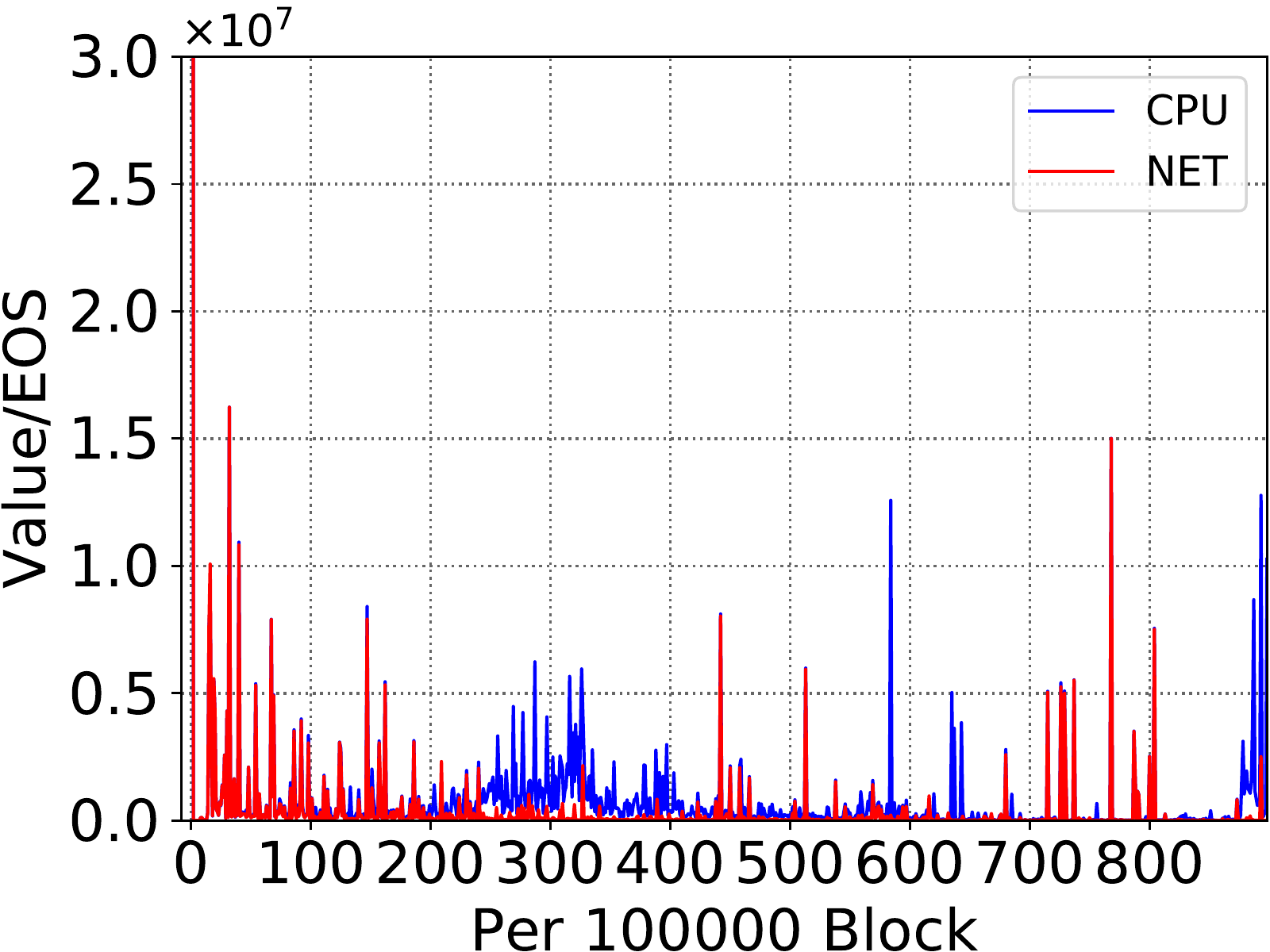} \label{7_stake}
}
\subfigure[\scriptsize Unstaked CPU and NET Amount]{
\centering
\includegraphics[width=5.0cm]{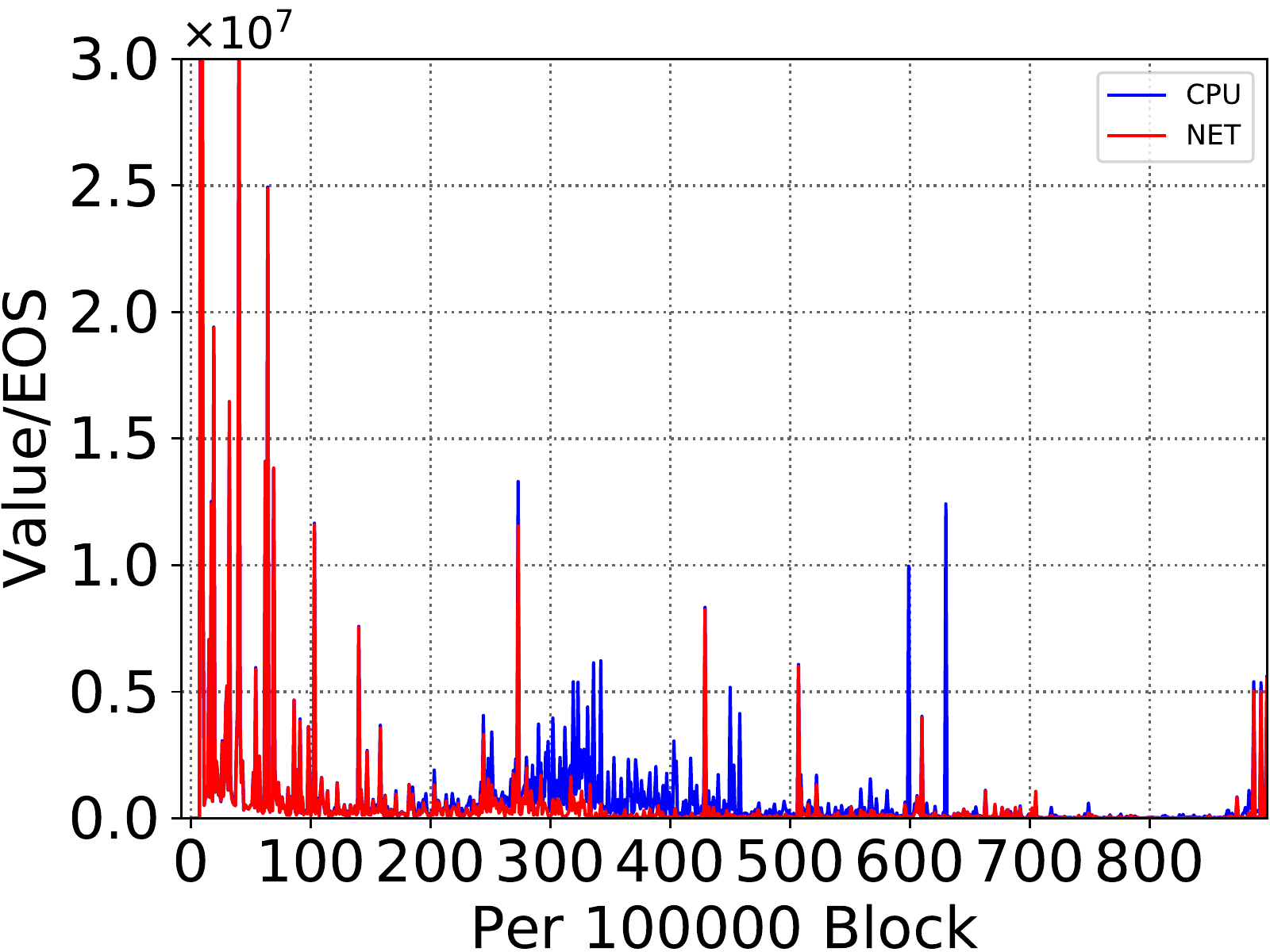} \label{7_unstake}
}
\subfigure[\scriptsize RAM-related Action Count]{
\centering
\includegraphics[width=5.0cm]{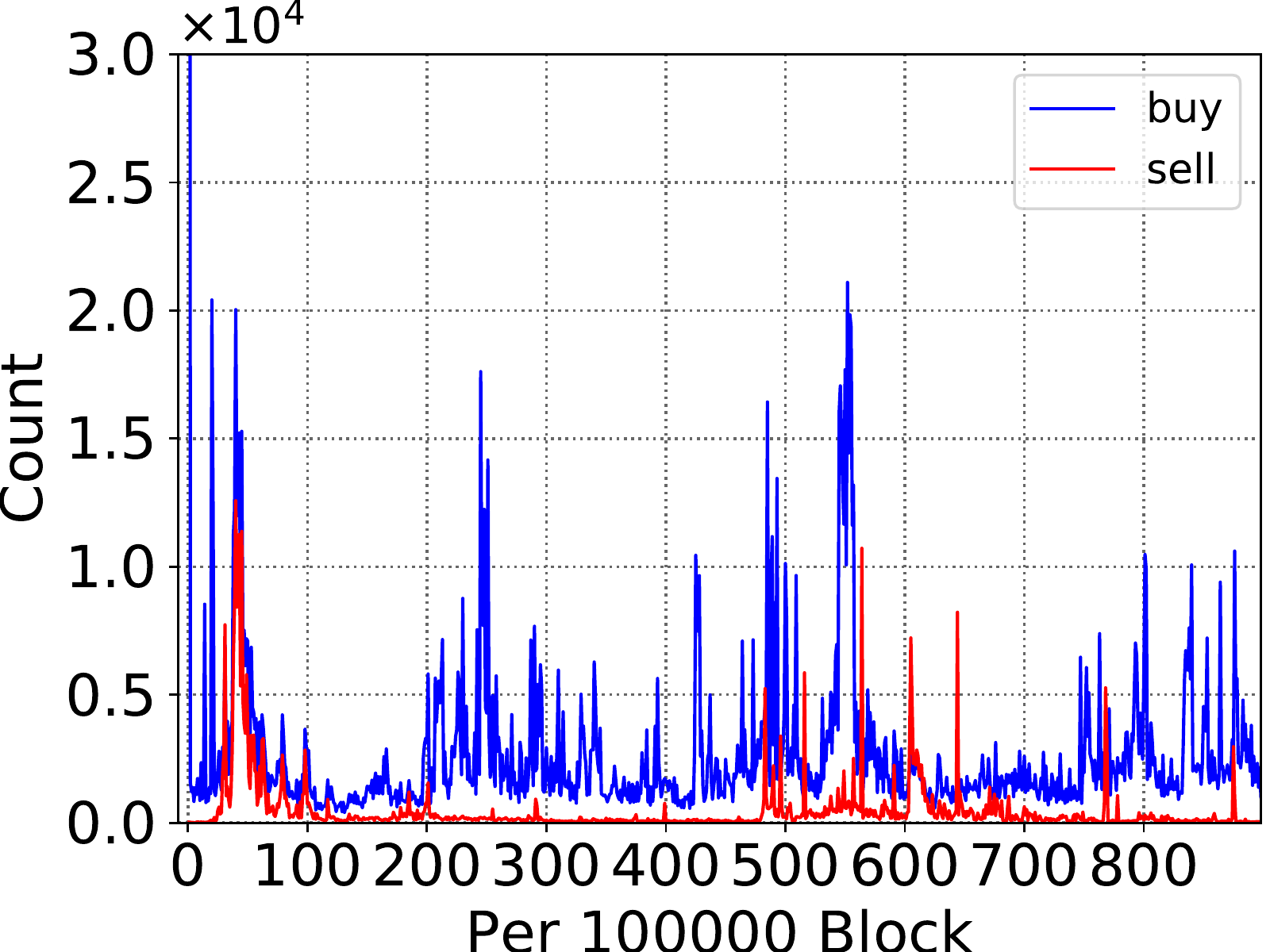} \label{7_ramcount}
}
\subfigure[\scriptsize RAM-related Action Amount]{
\centering
\includegraphics[width=5.0cm]{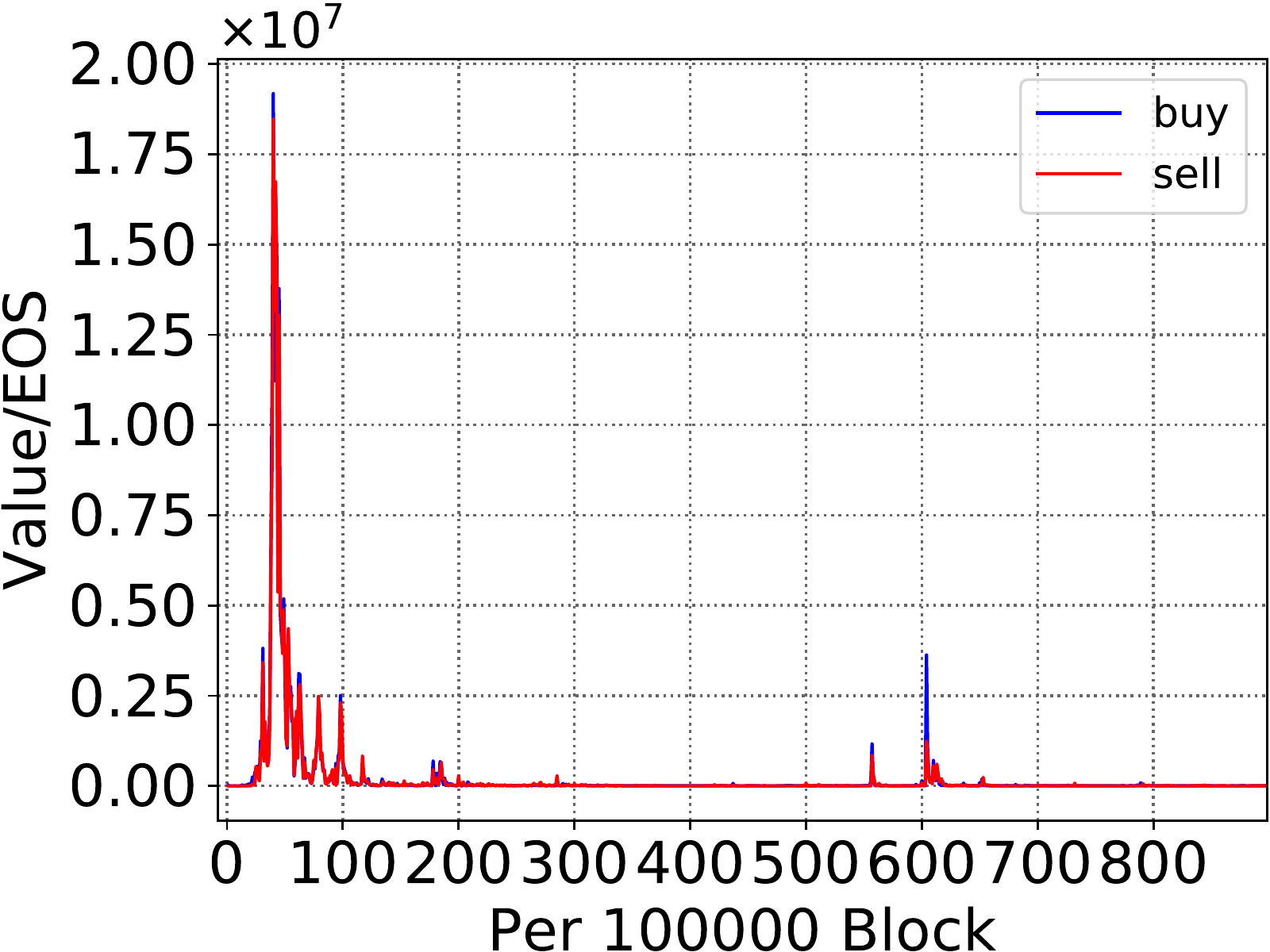} \label{7_ramamount}
}
\subfigure[\scriptsize REX Buy or Sell Amount]{
\centering
\includegraphics[width=5.0cm]{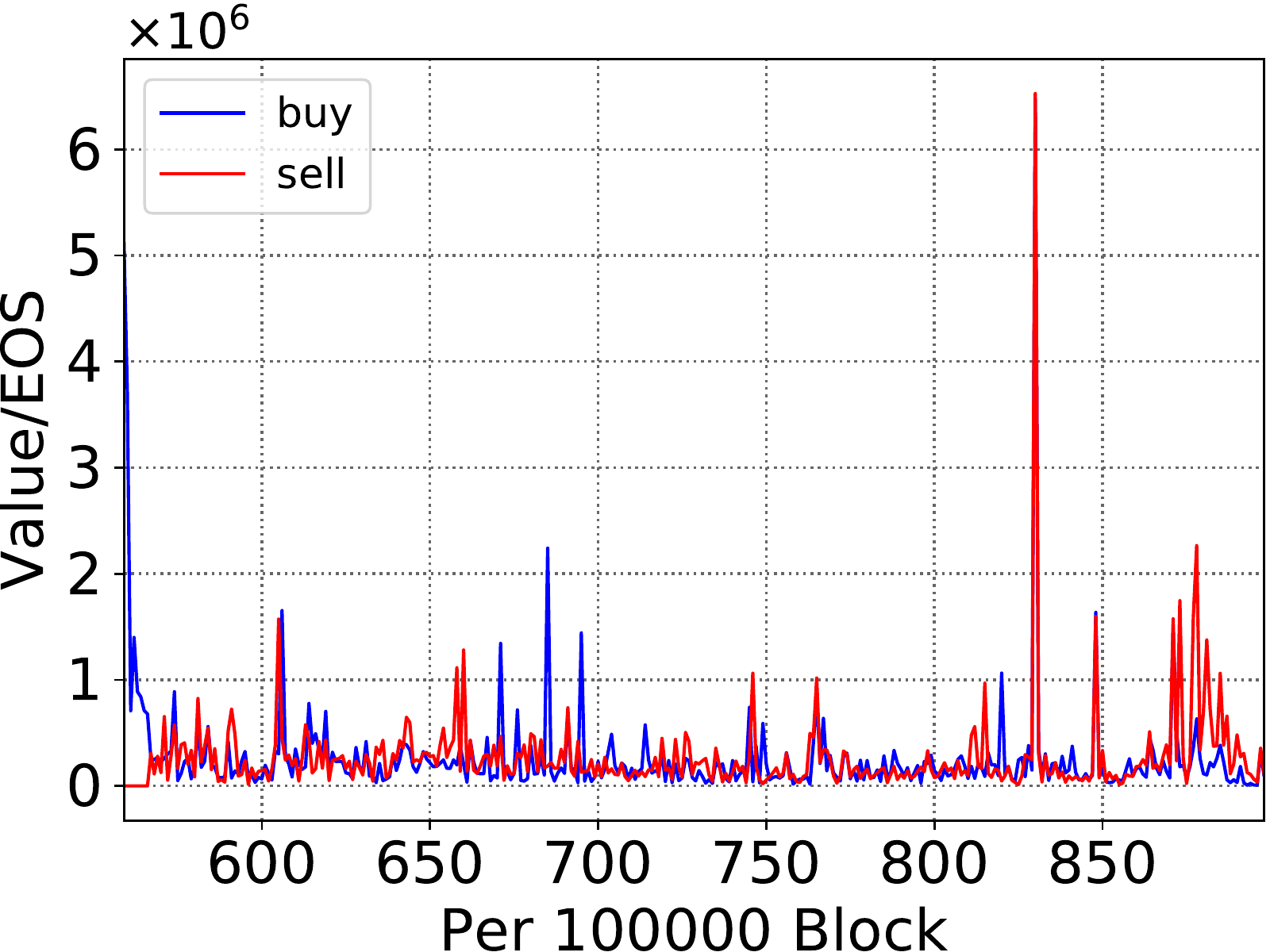} \label{7_rexamount}
}
\subfigure[\scriptsize CPU and NET Rent Amount]{
\centering
\includegraphics[width=5.0cm]{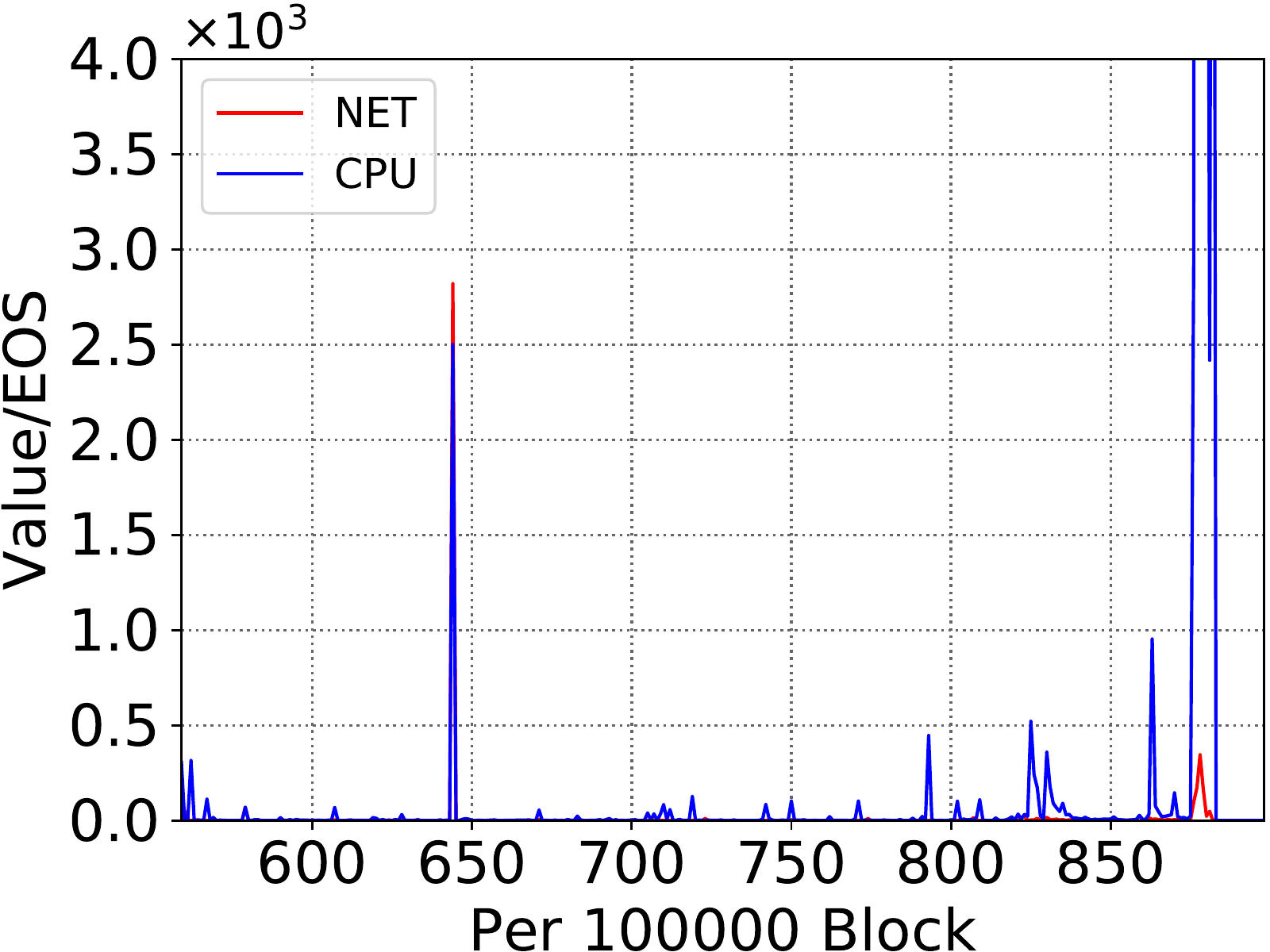} \label{7_rentamount}
}
\caption{Statistics of Dataset 7 (better viewed in color)}
\end{figure}

In order to investigate the resource management of EOSIO, we extract the actions related to CPU, NET, RAM, and REX from the raw data. As shown in Table~\ref{Statistics of Dataset 7}, there are 5,474,353 CPU-related actions, including 3,805,742 \texttt{\small stakecpu} actions and 1,668,611 \texttt{\small unstakecpu} actions. Meanwhile, there are 3,100,820 NET-related actions, including 2,324,444 \texttt{\small stakenet} actions and 776,376 \texttt{\small unstakenet} actions. Figures~\ref{7_stake} and~\ref{7_unstake} show the amount of EOS staked and unstaked, respectively. No matter being staked or unstaked, the amount of EOS corresponding to CPU is higher than that of NET, implying that CPU is a more ``\textit{important}'' resource compared with NET in EOSIO. In particular, when blocks reaching about 33,000,000, the amount of EOS staked surges because a large number of Gambling and Games DApps stake substantial CPU resource for users to gamble at this time.
\begin{table}[h]
    \setlength{\belowcaptionskip}{0.2cm}
    \centering
    \caption{Statistics of Dataset 7}
    \renewcommand{\arraystretch}{1.5}
    \footnotesize
    \begin{tabular}{|l|r|}
    \hline
    \textbf{Statistics} &  \textbf{Values} \\ \hline\hline
    No. of CPU-related Actions    &  5,474,353   \\ \hline    
    No. of NET-related Actions    & 3,100,820   \\ \hline    
    No. of RAM-related Actions    & 2,983,276   \\ \hline    
    No. of REX-related Actions    & 404,355   \\ \hline    
    \end{tabular}
    \label{Statistics of Dataset 7}
\end{table}

\renewcommand\subfigcapskip{-0.5ex}
% CPU account 3805742+1668611
% NET account 2324444+776376
% RAM account 2546849+436427
% REX 127318+51942+211075+14020

Since the launch of the EOSIO \textit{mainnet}, the speculation in RAM prices has continued. Some users hoarded RAM at low prices and sold at high prices to earn the difference profits. As shown in Table~\ref{Statistics of Dataset 7}, there are a total number of 2,983,276 RAM-related actions, including 2,546,849 \texttt{\small buyram} actions and 436,427 \texttt{\small sellram} actions. As shown in Figure~\ref{7_ramcount}, the count of \texttt{\small buyram} actions of every 100,000 blocks is significantly larger than that of \texttt{\small sellram}. Meanwhile, there are multiple peaks in both \texttt{\small buyram} and \texttt{\small sellram}. However, most of the time, there is not much difference between the EOS amount of \texttt{\small buyram} and that of \texttt{\small sellram} of every 100,000 blocks. It implies that users may buy RAM multiple times and sell it at once.

In order to solve the problem that users do not have enough EOS to stake CPU, EOSIO officially launched the CPU/NET leasing mechanism, i.e., the REX mechanism, on May 1, 2019 (around the block 56,000,000). Users can store some EOS tokens in REX pool through \texttt{\small buyrex} action to lease to others, and retrieve EOS and get the corresponding rent at any time through \texttt{\small sellrex} action. Meanwhile, users can rent CPU or NET from the REX pool by \texttt{\small rentcpu} or \texttt{\small rentnet} actions. In EOSIO, the rental income of the lessor will be affected by the supply-and-demand relationship between the lessor and lessee, so there is also speculation. As shown in Table~\ref{Statistics of Dataset 7}, there are 404,355 REX-related actions, including 127,318 \texttt{\small buyrex} actions, 51,942 \textit{sellrex} actions, 211,075 \texttt{\small rentcpu} actions, and 14,020 \texttt{\small rentnet} actions. Figure~\ref{7_rexamount} shows the EOS amount of \texttt{\small buyrex} and \texttt{\small sellrex} actions of every 100,000 blocks. Around the block 87,000,000, the EOS amount of both \texttt{\small buyrex} and \texttt{\small sellrex} are as high as 6 million EOS; it indicates that a large amount of funds enter and exit the REX pool at this time. In addition, from Figure~\ref{7_rentamount}, we can see that the EOS amount of both \texttt{\small rentcpu} and \texttt{\small rentnet} actions increase sharply around the block 64,400,000. For most of the time, the EOS amount of \texttt{\small rentcpu} actions is larger than that of \texttt{\small rentnet} actions, implying that users have higher demands for CPU.

\section{Applications of XBlock-EOS}
\label{sec:application}
\begin{figure}
    \centering
    \includegraphics[width=4.8in]{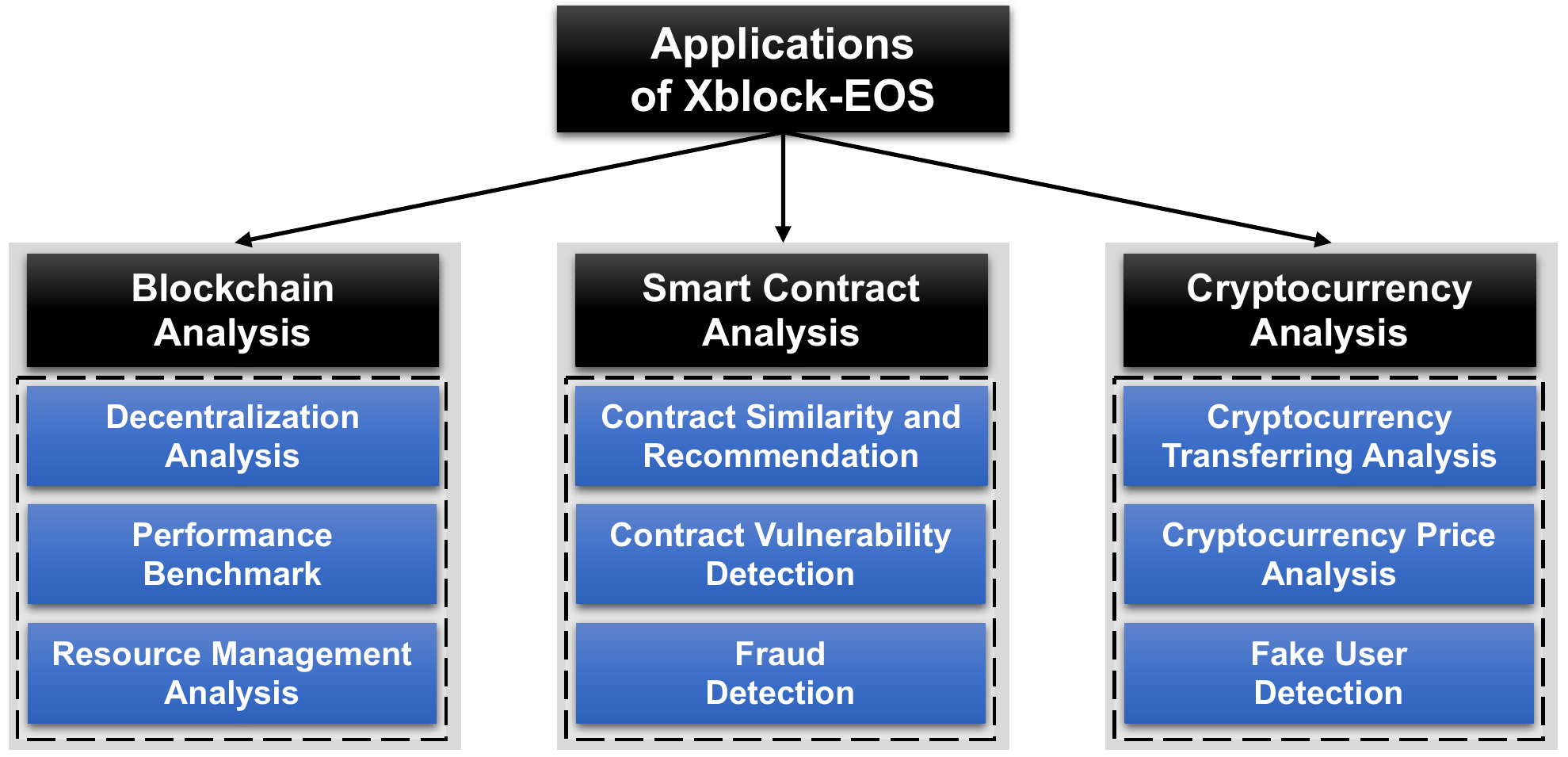}
    \caption{Applications of XBlock-EOS}
    \label{5_application}
\vspace*{-0.5cm}
\end{figure}

XBlock-EOS framework can support a diversity of emerging applications. As shown in Figure~\ref{5_application}, we categorize the applications of XBlock-EOS framework into three types: 1) Blockchain analysis in Section~\ref{subsec:blockchain}, 2) Smart contract analysis in Section~\ref{subsec:smart_contract}, and 3) Cryptocurrency analysis in Section~\ref{subsec:cryptocurrency}. This categorization is mainly based on the top-3 layers in the EOSIO architecture. Meanwhile, we also discuss the research opportunities in each layer, all of which contribute to building a more secure and healthier blockchain system. For example, as the key feature of blockchain, the decentralization analysis helps us to evaluate the security of a blockchain system. Meanwhile, resource management analysis helps us gain insight into the resource usage of the blockchain network and explore a more robust resource mechanism to avoid potential attacks (such as DDoS attacks). In addition, contract vulnerability detection is conducive to the security of contract funds, thereby preventing unexpected capital losses.

\subsection{Blockchain analysis}\label{subsec:blockchain}
XBlock-EOS that has processed a large amount of data from EOSIO blockchain system can be used to support the following applications for real-world users.

\subsubsection{Decentralization analysis}
Decentralization, as one of blockchain key characteristics, is also the core that brings numerous merits. However, there are few studies on the decentralization evaluation of different blockchain systems, especially for DPoS-based EOSIO. In the study \cite{wang2019measurement}, Wang et al. proposed a measurement study for the Bitcoin mining pool. Besides, Gencer et al. proposed a measurement study on the decentralization level of the Bitcoin and Ethereum networks~\cite{gencer2018decentralization}. These studies mainly focus on PoW-based blockchain systems and only consider few metrics such as network bandwidth, network structure, and mining power. In contrast, XBlock-EOS is responsible for the provision of massive processed data of \emph{DPOS-based blockchains} with multiple decentralization metrics. In addition, XBlock-EOS datasets can also be used to analyze the decentralization of users, contract owners, and block producers. Furthermore, the decentralization analysis on XBlock-EOS datasets can also be used to compare with blockchain systems (such as Bitcoin and Ethereum) with other consensus mechanisms like PoW.

\subsubsection{Performance benchmark}
Performance, especially for the transaction throughput is crucial to blockchains, thereby attracting extensive attention recently. There are many studies on blockchain performance optimization, such as Omniledger~\cite{kokoris2018omniledger}, RapidChain~\cite{zamani2018rapidchain}, and Monoxide~\cite{wang2019monoxide}. These studies often require a substantial amount of transaction data to evaluate the performance improvement. For example, Monoxide used historical Ethereum transaction data to evaluate the performance of its optimization scheme~\cite{wang2019monoxide}. In order to compare the performance of different optimization schemes fairly, a common benchmark for the real-world user cases for blockchain is needed although Zheng et al.~\cite{zheng2018detailed}, Xu et al.~\cite{XU2021102436} and BlockBench~\cite{dinh2017blockbench} proposed preliminary performance evaluation frameworks for different types of blockchain systems. More importantly, these performance evaluation frameworks need to simulate user behaviors and generate data similar to real-world blockchain systems. For example, Xu et al. use Hyperledger Caliper to generate transaction workloads, and perform experimental measurements to verify the correctness of their model~\cite{XU2021102436}. In this regard, the large amount of data in the XBlock-EOS framework can be regarded as an effective benchmark because XBlock-EOS data has been generated by users in the real world.

\subsubsection{Resource management analysis}
Unlike most public blockchains using the gas mechanism, EOSIO prevents the malicious behavior of smart contracts by limiting CPU, NET, and RAM resources. In EOISO, the total amount of these three resources is fixed so users (or contracts) need to compete for these resources to meet their own demands. Therefore, since the EOSIO \textit{mainnet} went live, speculation in resource price has continued, especially for CPU and RAM. We can learn from Section~\ref{Dataset 7: Resource Management} that the count of actions related to RAM and CPU fluctuates sharply at some moments. In addition, the launch of the REX mechanism has also caused a speculative boom. Analyzing the user behaviors on resource management and identifying some speculative models in EOSIO can help users predict the price of resources and stagger the peak of speculation, thereby saving money for buying or staking resources. At present, there are some studies on the resource (gas) mechanism such as~\cite{chen2017adaptive, pierro2019influence} while few of them focus on the EOSIO's resource management model.

In addition, the security of the blockchain resource management model is also a hot topic that attracts much attention. For example, it is reported in 2016 that the Ethereum \textit{mainnet} suffered from a large-scale DDoS attack, which severely blocked the entire network. The attackers leveraged the vulnerability of inappropriate setting \textit{EXTCODESIZE} instruction in the gas mechanism to launch this DDoS attack. Therefore, the security of the resource management model is extremely important for blockchain systems. Currently, there are a few studies on the security of gas mechanism. For example, Chen et al. proposed an adaptive gas cost mechanism for Ethereum to defend against under-priced DoS attacks~\cite{chen2017adaptive}. Meanwhile, Lee et al. reported some threads to the new EOSIO's resource management model and proposed some mitigation methods~\cite{lee2019spent}. The analysis on the new EOSIO's management resource model can be compared with those of other blockchain systems (e.g., the gas mechanism) to promote the development of blockchain in this aspect. In summary, the resource management analysis in EOSIO will not only bring huge economic value but also promote the in-depth technology mature of blockchain systems, especially in security.

\subsection{Smart contract analysis}\label{subsec:smart_contract}
Similar to Ethereum, EOSIO also supports smart contracts. Priory analysis presented in Section~\ref{Dataset 4: Contract Invocation} shows that when the network is active, the count of contract invocations nearly reaches 6,000,000 per 100,000 blocks, i.e., near 120 invocations per second. It implies massive smart contracts in EOSIO are in an active state and the analysis of EOSIO smart contracts is worthwhile for conducting in-depth research in the future. We summarize the potential applications of XBlock-EOS on smart contracts as follows.

\subsubsection{Contract similarity and recommendation}
From the analysis in Sections~\ref{Dataset 3: Contract Information} and~\ref{Dataset 4: Contract Invocation}, we observe that there is a great similarity between \emph{smart contracts} and \emph{contract invocations}. Code similarity evaluation and detection, especially similarity detection have been a traditional research topic in software engineering~\cite{chilowicz2009syntax, luo2014semantics}. Recently, some studies have also been focused on the similarity analysis of smart contract. Norvill et al. proposed a framework to group together similar contracts within the Ethereum network, and further automate labeling unknown contracts~\cite{norvill2017automated}. He et al. performed a detailed similarity comparison of a large number of contracts to investigate the correlation between code reuse and vulnerabilities~\cite{he2019characterizing}. In addition, finding similar contracts is helpful to develop new contracts. For example, developers can estimate the effects of new contracts by estimating user behaviors before deploying a new contract. In the study~\cite{huang2019recommending}, Huang et al. proposed a method based on existing smart contracts to recommend distinguishing codes for updating smart contracts, which can help developers improve the code. Similarly, in terms of users, recommending similar smart contracts will help users find suitable contracts for them.

\subsubsection{Contract vulnerability detection} \label{Contract Vulnerability Detection}
Smart contract security has been a hot research topic in the blockchains, especially in Ethereum, EOSIO, and other blockchain systems. A series of attacks in Ethereum, such as TheDAO attack, have caused huge economic losses~\cite{mehar2019understanding}. Since EOSIO went online, a number of vulnerabilities have been discovered from EOSIO's smart contracts, including fake EOS transfer, fake transfer notice and flawed random numbers generators~\cite{quan2019evulhunter}. These vulnerabilities have also brought huge economic losses, especially for Gambling and Games DApps. In order to prevent malicious attacks on smart contracts, it is an important step to perform vulnerability detection before launching online. Recently, many studies have focused on the vulnerability detection of smart contracts in Ethereum, such as Oyente~\cite{luu2016making} and Zeus~\cite{kalra2018zeus}. Meanwhile, a few studies have performed similar analysis on EOSIO smart contracts. For example, EVulHunter can detect fake transfer vulnerabilities for EOSIO's Smart Contracts at Webassembly-level~\cite{quan2019evulhunter}. In some cases, the vulnerability detection of smart contracts may be inspired by traditional software vulnerability detection approaches, which can also verify the code. Some studies have focused on verifying the contract codes (also called ``\textit{bytecodes}'' or ``\textit{opcodes}''). In this respect, the contract code data collected by XBlock-EOS can also be applied to contract vulnerability detection.

\subsubsection{Fraud detection}
The rapid development and prosperous popularization of smart contracts bring huge economic values. Meanwhile, smart contracts have become a means employed by malicious users to design scams to make exorbitant profits. For example, crowdfunding contracts ostensibly bring a promised return to attract victims to invest. The study~\cite{Weili18} shows that the Ponzi scam can deceive others' cryptocurrencies. Currently, a few studies have proposed several methods for detecting fraudulent contracts and corresponding fraud activities in Ethereum~\cite{chen2019exploiting,Torres:2019}. In addition, the study~\cite{huang2020characterizing} gain some insights that some (contract) accounts in EOSIO are \emph{bot-like} and can be used for malicious and fraudulent purposes including Bonus Hunting, Clicking Fraud, etc. Fortunately, most of these studies are based on the analysis of money transfers, contract codes, and contract invocations, all of which have been contained in XBlock-EOS. Therefore, XBlock-EOS can also support further research on fraud detection.

\subsection{Cryptocurrency analysis}\label{subsec:cryptocurrency}
Since the ICO waves in 2017, blockchain-based cryptocurrencies have received much attention due to decentralization and cost reduction. Ethereum, EOSIO and other blockchain systems contain a large number of cryptocurrencies including native cryptocurrencies such as ETH and EOS, as well as tokens issued in accordance with a certain protocol. For example, many tokens are issued in EOSIO using the contract \textit{eosio.token}. It is shown in CoinMarketCap\footnote{\url{https://coinmarketcap.com/all/views/all/}} that until now, there are more than 5,000 tokens used for third-party exchange with a market value of up to 15 billion dollars. Therefore, analyzing the cryptocurrency data on the blockchain can bring great economic value. We summarize the potential applications of XBlock-EOS on cryptocurrency, as described below.

\subsubsection{Cryptocurrency transferring analysis}
Cryptocurrency transferring analysis is common in cryptocurrency analysis. Chen et al.~\cite{chen2018understanding} conducted a graphical analysis of Ether transfers and derive some interesting insights. Moreover, Victor et al.~\cite{victor2019measuring} and Chen et al.~\cite{tokenscope} analyzed the ERC20/ERC721 token transfers network in Ethereum. Most recently, Huang et al.~\cite{huang2020characterizing} proposed the graph analysis on EOS transfers to assist in detecting some bots and fraudulent activities in EOSIO. Following the analysis of cryptocurrency transfers, the further analysis on user behaviors can be done. For example, different tokens may form different user communities. 

In addition, due to the anonymity of cryptocurrencies, blockchain has become a tool for money laundering. Cryptocurrency transferring analysis will help to identify and detect money laundering behaviors on the blockchain, consequently promoting the development of blockchain regulation. The massive processed data of XBlock-EOS will further facilitate the research on the above issues.

\subsubsection{Cryptocurrency price analysis}
The price of blockchain-based cryptocurrencies is susceptible to a number of different factors, such as government policies,  technological innovation, socioeconomic status and related business activities. The price of cryptocurrencies is of greatest concern to the general public. Some recent studies have focused on the analysis and prediction of cryptocurrency prices~\cite{lamon2017cryptocurrency, abraham2018cryptocurrency, mensi2019structural}. These studies generally consist of three steps: (1) collect the price data of cryptocurrency from centralized third-party exchanges, (2) analyze the correlation between cryptocurrency prices and other potential factors, (3) forecast the future prices and potential profits. The data extracted by our XBlock-EOS can be used for the analysis to evaluate potential factors affecting the prices of cryptocurrencies, especially in the second step analysis, which is the most critical step. In addition, EOSIO data collected by XBlock-EOS collects can be jointly analyzed with other datasets collected from other blockchain systems (such as Ethereum~\cite{zheng2019xblock}) to investigate the prices of cryptocurrencies and even the correlation between different cryptocurrency prices from multiple perspectives.

\subsubsection{Fake user detection}
Fake user detection has always been a hot research topic in social networks~\cite{cao2012aiding, varol2017online}, which can help to avoid the economic loss caused by malicious activities. The cryptocurrency users in blockchain systems also form a community-like social network, in which fake users controlled by some developers can falsify clicks to improve the rankings of DApps and even launch malicious activities. Generally speaking, the rankings of DApps or cryptocurrencies are mainly based on some objective indicators related to user activities, such as daily active users and daily user transaction volumes. Malicious developers may employ this mechanism to create fake users through various methods to increase the activity rankings of DApps. The study~\cite{huang2020characterizing} shows that more than 30\% of accounts in EOSIO were bot-like accounts through graph analysis. In addition, DAppReview\footnote{\url{http://dapp.review}} labels cryptocurrencies with fake users, but this fake user detection method is mainly conducted in a black box and requires manual operations. At present, there are few studies on fake user detection on DApps or cryptocurrencies. Compared to permissioned blockchain systems, there may be more fake account activities on permissionless blockchain systems. Fake user detection is of great significance for the healthy development of the permissionless blockchains. In this regard, our XBlock-EOS can support the further study on fake user detection so as to establish healthier blockchains.

\section{Related work and discussion} 
\label{sec:related}
In this section, we introduce and discuss some existing studies on blockchain data analysis. We categorize the state-of-the-art literature into two categories: \textit{Data collection} and \textit{Data analysis}.

\emph{Data collection} is the prerequisite for blockchain data analysis while it is often challenging due to the massive volumes of blockchains. At present, the studies on blockchain data collection are mainly focused on Bitcoin and Ethereum while few of them are concentrated on EOISO. For example, DataEther~\cite{chen2019dataether} is a tool with code modification of the Ethereum clients to obtain the Ethereum data, while Google BigQuery~\cite{tigani2014google} imports data from Bitcoin and Ethereum to enable researchers to analyze data online. XBlock-ETH~\cite{zheng2019xblock} provides a large number of well-processed Ethereum datasets while does not include the EOSIO data. Regarding EOSIO's data collection tools, some blockchain data browsers provide data APIs for developers to use and analyze. For example, \textit{EOSPark}\footnote{\url{https://eospark.com/}} provides a web interface to support comprehensive queries on EOSIO data as well as some simple analysis tools on EOSIO data including blocks, transactions, contracts and tokens. In addition, \textit{eosq}\footnote{\url{https://eosq.app/}} provides high-precision queries on EOSIO data, supporting the queries on the detailed calling information of each transaction. However, these third-party blockchain data service providers have a number of limitations on user permissions and data usage. It is impossible for researchers to collect all the EOSIO data through these platforms. In short, most of the above data tools only provide users with tools or API services while they do not provide the up-to-date well-processed datasets.

\emph{Data analysis} of blockchain data has mainly focused on Ethereum and Bitcoin, especially Ethereum. The studies on Ethereum data analysis mainly include transaction analysis, fraud detection, smart contract security, and token analysis~\cite{Weili18,chen2019exploiting,chen2018understanding,tokenscope, kalra2018zeus, luu2016making}. Compared with Ethereum, EOSIO has higher volumes of various types of blockchain data. However, there are relatively few studies on the EOSIO data. Huang et al. characterize the activities in EOSIO including money transfers, account creation and contracts to detect bots and fraudulent activities~\cite{huang2020characterizing}. In addition, EVulHunter~\cite{quan2019evulhunter} presents the first systematic attempt to automatically detect fake-transfer vulnerabilities of EOSIO's smart contract at Webassembly-level. Although these studies publish some specific types of EOSIO data, they can only be applicable to specific analysis. Moreover, datasets released by these studies lack maintenance and update.

To our best knowledge, XBlock-EOS is the first to provide such comprehensive EOSIO raw data as well as well-processed datasets. All datasets in our XBlock-EOS are public and conveniently accessible to various data users (from data analysts to DApp developers). The analysis on these data can bring huge economic values and promote the further benign development of EOSIO blockchain. Moreover, XBlock-EOS keeps updating the datasets regularly to maintain the latest EOSIO data being synchronized with the EOSIO \textit{mainnet}.

\section{Conclusion and future work} \label{sec:conc}
In this paper, we introduce a data collection framework of EOSIO data namely XBlock-EOS, which contains a well-processed up-to-date on-chain data of EOSIO, including blocks, transactions, actions, contracts, tokens, accounts, and resources. Moreover, this paper also presents comprehensive statistics and exploration of these processed datasets. We also discuss the emerging applications based on XBlock-EOS and outline future research opportunities. Currently, the XBlock-EOS datasets have been published on the \textbf{\textit{XBLOCK.PRO}} website\footnote{\url{http://xblock.pro}\label{website}}, through which every user can easily obtain them.

Our XBlock-EOS is promising to promote the studies in EOSIO and advance the development of blockchains. The future improvements of XBlock-EOS are described as follows:

\textbf{(1) Collect off-chain data from exchanges and open-source communities}: Off-chain data is also very important for blockchain data analysis, as it provides off-chain behavior information for blockchain users and developers. Our XBlock-EOS will offer the off-chain data in the future.

\textbf{(2) Explore more features}: This paper introduces the basic characteristics of the EOSIO data. Compared to Ethereum, EOSIO can be considered as a completely novel public blockchain system. EOSIO's architecture and design principles are very different from those of Ethereum. In the future, we will further explore the features of EOSIO data.

\textbf{(3) Perform a combination analysis with other blockchain systems:} In recent years, the rapid development of blockchain technologies as well as the prosperous blockchain applications have attracted a large number of users and developers. The joint analysis of EOSIO with other blockchain systems will be conducted in the future.

% or


% \appendices

% % you can choose not to have a title for an appendix
% % if you want by leaving the argument blank
% \section{}

\section*{Acknowledgements}
The work described in this paper was supported by the National Key Research and Development Program (2016YFB1000101), the National Natural Science Foundation of China (61722214), Key-Area Research and Development Program of Guangdong Province (2019B020214006) and Macao Science and Technology Development Fund under Macao Funding Scheme for Key R \& D Projects (0025/2019/AKP). %Zibin Zheng is the corresponding author.

% \section*{References}

\bibliography{main.bib}

\end{document}